\documentclass[sigconf]{acmart}
\usepackage{etex}

\usepackage{algorithm,algorithmicx,algpseudocode}
\usepackage{amstext}   
\usepackage{amsmath}
\usepackage{amssymb}
\usepackage{lipsum}
\usepackage{float} 
\usepackage{balance}
\usepackage{caption}
\usepackage{color}
\usepackage{comment}
\usepackage{enumitem}
\usepackage{epic}
\usepackage{epsf}
\usepackage{epsfig}
\usepackage[T1]{fontenc}
\usepackage{graphicx}
\usepackage{hyperref}
\usepackage{latexsym}
\usepackage{listings}
\usepackage{makecell}
\usepackage{multirow}
\usepackage{multicol}
\usepackage{pifont}
\usepackage{soul}
\usepackage[subrefformat=parens,labelformat=parens]{subfig}
\usepackage{textcomp}
\usepackage{tabularx}
\usepackage{url}
\usepackage{xspace}    
\usepackage{tikz}
\usepackage[color=white]{todonotes}
\usepackage[utf8]{inputenc} 

\newcommand{\circlenum}[1]{\tikz \draw[fill=black,text=white,inner sep=0,outer sep=0,baseline=O.base,font=\fontsize{8}{\baselineskip}\selectfont] (0,0) circle (0.16cm) node (O) {#1};\ }

\newcommand{\circledash}[1]{\tikz \draw[fill=white,white,text=black,inner sep=0,outer sep=0,baseline=O.base,font=\fontsize{8}{\baselineskip}\selectfont] (0,0) circle (0.12cm) node (O) {-};}

\lstdefinestyle{codesnip}{
  numbers=left,
  numbersep=-5pt,                  
  language=C,
  basicstyle=\linespread{0.1}\scriptsize\ttfamily,
  escapechar=\&
}
\newfloat{lstfloat}{htbp}{lop}
\floatname{lstfloat}{Listing}

\graphicspath{{figs/}}

\newcommand{\ignoreme}[1]{}

\newcommand{\papernamen}{FirmUSB\xspace}
\newcommand{\papername}{\textsc{\papernamen}\xspace}
\newcommand{\inprogress}[1]{\grant{This section is in progress}} %
\newcommand{\outline}[1]{}

\newcommand{\grant}[1]{}

\newcommand{\angr}{\textsc{angr}\xspace}
\newcommand{\fie}{\textsc{Fie}\xspace}
\newcommand{\klee}{\textsc{KLEE}\xspace}
\newcommand{\ste}{S$^{2}$E\space}
\newcommand{\backendangr}{\angr Engine\xspace}
\newcommand{\backendfie}{\fie Engine\xspace}

\makeatletter
\def\blfootnote{\xdef\@thefnmark{}\@footnotetext}
\makeatother

\makeatletter
\newcommand\footnoteref[1]{\protected@xdef\@thefnmark{\ref{#1}}\@footnotemark}
\makeatother

\makeatletter
\newcommand{\labitem}[2]{%
\def\@itemlabel{\textbf{#1}}
\item
\def\@currentlabel{#1}\label{#2}}
\makeatother

\title{\textsc{\papernamen}: Vetting USB Device Firmware using Domain Informed Symbolic Execution}

\author{Grant Hernandez}
\affiliation{%
  \institution{University of Florida}
  \city{Gainesville}
  \state{FL}
  \country{USA}
}
\email{grant.hernandez@ufl.edu}
\authornote{These authors have contributed equally to this work.}

\author{Farhaan Fowze}
\affiliation{%
  \institution{University of Florida}
  \city{Gainesville}
  \state{FL}
  \country{USA}
}
\email{farhaan104@ufl.edu}
\authornotemark[1]

\author{Dave (Jing) Tian}
\affiliation{%
  \institution{University of Florida}
  \city{Gainesville}
  \state{FL}
  \country{USA}
}
\email{daveti@ufl.edu}

\author{Tuba Yavuz}
\affiliation{%
  \institution{University of Florida}
  \city{Gainesville}
  \state{FL}
  \country{USA}
}
\email{tuba@ece.ufl.edu}

\author{Kevin R. B. Butler}
\affiliation{%
  \institution{University of Florida}
  \city{Gainesville}
  \state{FL}
  \country{USA}
}
\email{butler@ufl.edu}

\renewcommand{\shortauthors}{G. Hernandez and F. Fowze et al.}

\begin{document}

\begin{abstract}
The USB protocol has become ubiquitous, supporting devices from high-powered
computing devices to small embedded devices and control systems.
USB's greatest feature, its openness and
expandability, is also its weakness, and attacks such as BadUSB exploit the
unconstrained functionality afforded to these devices as a vector for
compromise. Fundamentally, it is virtually impossible to know whether a
USB device is benign or malicious. This work introduces \papername, a
USB-specific firmware analysis framework that uses domain
knowledge of the USB protocol to examine firmware images and determine the
activity that they can produce. Embedded USB devices use microcontrollers
that have not been well studied by the binary analysis community, and our
work demonstrates how lifters into popular intermediate representations for
analysis can be built, as well as the challenges of doing so. We develop
targeting algorithms and use domain knowledge to speed up these processes by
a factor of 7 compared to unconstrained fully symbolic execution.
We also successfully find malicious activity in
embedded 8051 firmwares without the use of source code. Finally, we provide
insights into the challenges of symbolic analysis on embedded architectures
and provide guidance on improving tools to better handle this important
class of devices.

\end{abstract}

\copyrightyear{2017}
\acmYear{2017}
\setcopyright{acmcopyright}
\acmConference{CCS '17}{October 30-November 3, 2017}{Dallas, TX, USA}\acmPrice{15.00}\acmDOI{10.1145/3133956.3134050}
\acmISBN{978-1-4503-4946-8/17/10}




\maketitle

\section{Introduction}

The Universal Serial Bus (USB) protocol enables devices to communicate with each
other across a common physical medium.  USB has become 
ubiquitous and is supported by a vast array of devices, from 
smartphones to desktop PCs,  small peripherals, such as
flash drives, webcams, or keyboards, and even control systems and other
devices that do not present themselves as traditional computing platforms.
This ubiquity allows for easy connecting of devices to data and power.
However, attacks that exploit USB have become increasingly common and
serious. As an example the {\em BadUSB} attack exploits the open nature of the
USB protocol, allowing the advertisement of capabilities that device users
may not realize are present. A BadUSB device appears to be a benign flash
drive, but advertises itself as having keyboard functionality when plugged
into a victim's computer; the host 
unquestioningly allows such a capability to be used. The malicious device is
then able to inject keystrokes to the computer in order to bring up a
terminal and gain administrative access. Fundamentally, there is an
inability to constrain device functionality within USB, coupled with a 
corresponding lack of ability to know what types of functionalities a device is
capable of advertising and whether or not these are benign.

Previous work has focused on preventing USB attacks at the protocol level, through isolation-based approaches such as sandboxing and
virtualization~\cite{angel2015defending,tsb+16} or involving the user in the
authorization process~\cite{tbb15}. These
approaches suffer from a common problem: they rely on a device's external
actions to demonstrate its trustworthiness. Without a deeper understanding of the
underlying software controlling these devices, an external observer cannot with
certainty ensure that a device is trustworthy.
Even solutions such as signed
firmware give little evidence of its actual validity; signing merely
demonstrates that an entity has applied their private key to a firmware, but
does not in itself provide any assurance regarding device integrity.
Consequently, there is limited ability to validate the trustworthiness
and integrity of devices themselves.

In this paper, we address these concerns through the analysis of firmware
underlying USB devices.  We create
\papername, a framework that uses domain knowledge of the USB protocol to
validate device firmware against expected functionality through symbolic
execution.  USB devices are
often small and resource-constrained, with significantly different chip
architectures than the ARM and x86 processors found on computers and
smartphones. While substantial past work has focused on firmware analysis
of these processor
architectures~\cite{chen2016-wu,SWH15}, comparatively little has been done on the microcontrollers
that embedded USB devices often employ. 
We bring architecture-specific support to existing frameworks
and provide informed guidance through USB-specific knowledge to improve analysis. 
We have designed and implemented binary lifters to allow for symbolic
analysis of the Intel 8051 MCU, which represents a
Harvard architecture chip designed in 1980 that looks vastly different from
modern processor designs, but is commonly used in USB flash drives as well
as many other embedded environments. 
We use two symbolic execution frameworks for our analysis in order to better
understand the benefits and challenges of different approaches when using
uncommon instruction architectures. We use \fie~\cite{DMR13}, which uses LLVM as an
Intermediate Representation (IR) and is built on top of the popular \klee~symbolic
execution engine~\cite{CDE08}, as well as \angr~\cite{SWS16}, which is designed to be used
for binary analysis and employs Valgrind's VEX as an IR.
\papername is {\em bottom-up}, in that it does not rely on the existence
of source code to perform its analysis. This is crucial for microcontroller
firmware, for which source code may be difficult if not impossible to publicly find
for many proprietary USB controllers.
\papername uses static analysis and symbolic execution, to extract the
semantics of a firmware image in order to build a model of discovered firmware
functionality for comparison to expected functionality.

Our contributions are summarized as follows:
\begin{itemize}
    \item {\bf Firmware Analysis Framework:} We develop a USB-specific firmware analysis framework to verify or determine the intention
of compiled USB controller firmware binaries running on the 8051/52 architectures.
To our knowledge this is the first 8051 lifter into the popular VEX and
LLVM IRs.
\item {\bf Domain-Informed Targeting:} We show that \papername detects
    malicious activity in Phison firmware images for flash drive
    controllers containing BadUSB, as
    well as EzHID HID firmware images for 8051 containing malicious activity.
    For the malicious Phison image, our domain-specific approach speeds up
    targeting by a factor of 7 compared to unconstrained fully symbolic execution. 
\item {\bf Analysis of Existing Symbolic Frameworks:} We provide insights
    and describe the challenges of utilizing existing tools to analyze
    binary firmware for embedded systems architectures, and present
    guidance on how such tools can be improved to deal with these
    architectures.
\end{itemize}

\paragraph{Outline}
The rest of this paper is structured as follows: \autoref{sec:background}
provides background on embedded firmware analysis, our case study on the 8051
architecture in the context of USB devices, and our major challenges in
analyzing black-box firmware using symbolic execution. \autoref{sec:overview}
presents a high-level overview of \papername and \autoref{sec:design} follows
with low-level details. \autoref{sec:eval} evaluates the performance of our
\angr and \fie
implementations on crafted 8051/52 binaries. We discuss key takeaways
from our work in \autoref{sec:disc} and mention what difficulties we
experienced during development. We discuss related work in \autoref{sec:relwork} and conclude in \autoref{sec:conc}.

\section{Background}
\label{sec:background}
\subsection{Universal Serial Bus}

The USB protocol provides a foundation for host-peripheral communications
and is a ubiquitous interface. USB is a host-master protocol, 
which means that the host initiates all communication on the underlying
bus.\footnote{USB OTG and USB 3.0 are the exceptions. While USB 3.0 and later devices allow for device-initiated
communication, such a paradigm is still relatively rare amongst
peripherals, which are overwhelmingly designed to respond to host queries.}
This is true even for interrupt driven devices such as keyboards.  The
underlying bus arbitration and low-level bit stream are handled in dedicated
hardware for speed and reliability. In our work, we primarily focus on the
device level configuration and omit the study of lower-level aspects of USB
(i.e. power management, speed negotiation, timing).

When a USB device is first plugged in to a host
machine, undergoes the process of \emph{enumeration}. A device possesses a
set of {\em descriptors} including device, configuration, interface, and endpoint
descriptors. A \emph{device descriptor} contains the vendor (VID) and product (PID)
identifiers, pointers to string descriptors, and device class and protocol.
VIDs are assigned to the vendor by the USB Implementor's Forum (USB-IF).
Vendors are then able
to assign arbitrary PIDs to their products.  VIDs and PIDs should be unique but
are not required to be. The device class (\texttt{bDeviceClass}) and its defined protocol (\texttt{bDeviceProtocol}) hint to the
host what capabilities to expect from the device. The last field in the device descriptor is the number of configurations (\texttt{bNumConfigurations}).
A USB device may have multiple \emph{configuration descriptors},
but only one may be active at a time. This high level descriptor describes the
number of interfaces and power characteristics.  \emph{Interface descriptors} have a
specific interface class and subclass. This defines the expected command set to
the host operating system.

Two important device classes in the context of this
paper are the Human Interface Device (HID) (\texttt{0x03h}) and the Mass
Storage (\texttt{0x08h}) classes. Devices are free to have mixed-class
interfaces, which means they are considered to be composite devices. For
example, a physical flash drive could contain two interfaces -- one mass
storage and the other HID. This would allow it to transfer and store bulk data
while possibly acting as a keyboard on the host machine. Additionally, a device
could at runtime switch configurations from a pure mass storage device to a
HID device. The final descriptor of interest is the \emph{endpoint descriptor}.
Endpoints are essentially mail boxes that have a direction (in and out),
transfer type (control, isochronous, bulk, or interrupt), poll rate, and
maximum packet size. By default, devices' first endpoint (Endpoint0 or EP0)
respond to \emph{control transfers}, which are mainly configuration details and
commands from and to the host machine. Other endpoints may be used for pure
data transfer.

The elements of the USB protocol that are implemented in hardware
and firmware varies based on to the specific USB controller. For instance, some USB devices may be
completely fixed in hardware, meaning that their configuration descriptors,
including their vendor and product IDs, are static. In this work, we assume
that the firmware deals with all of the major descriptors and the hardware just
provides low-level USB signaling.

\begin{figure*}[t]
\centering
\includegraphics[width=12cm]{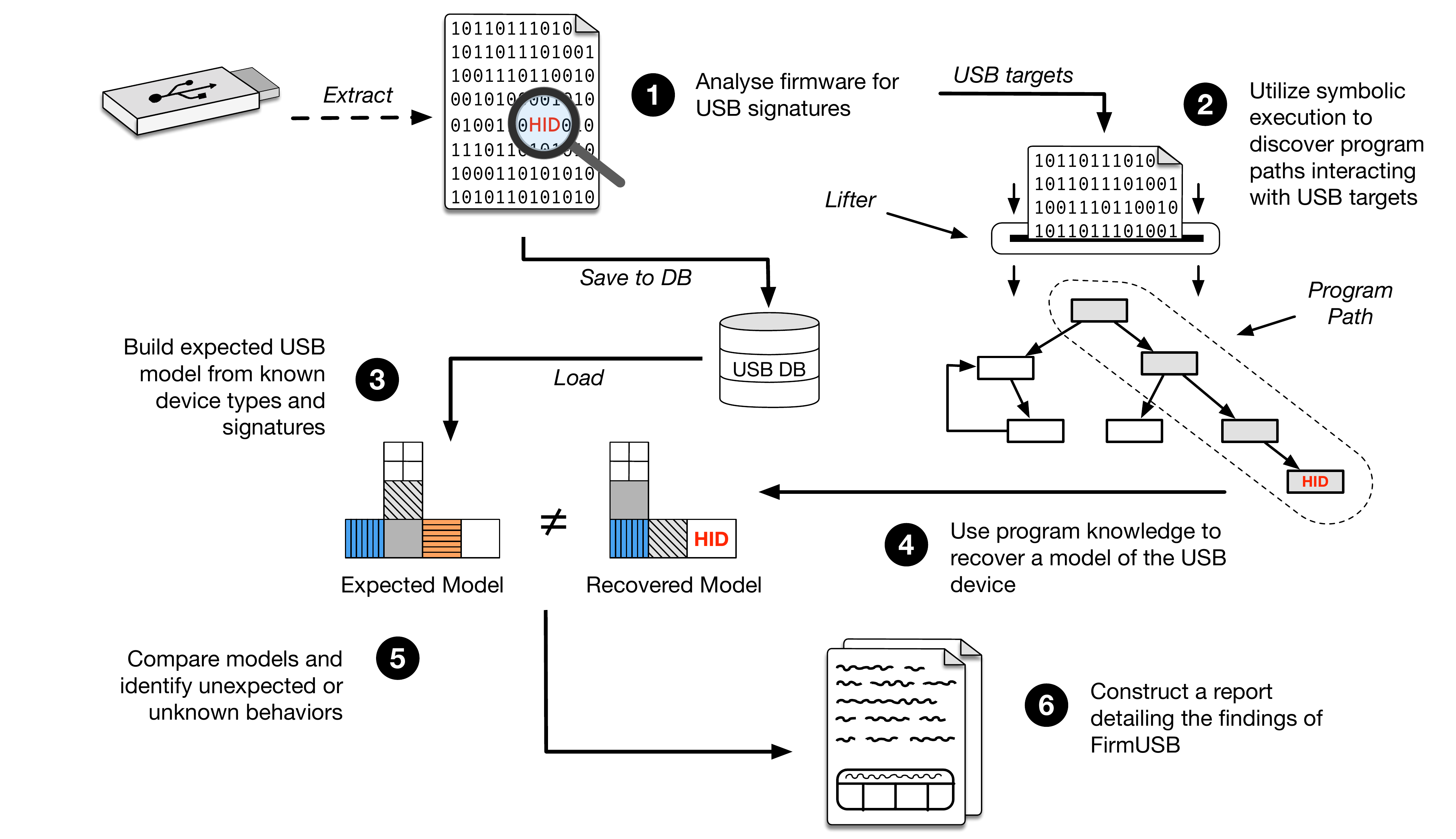}
\caption{An overview of \textsc{\papernamen}'s primary flow through analyzing firmware.}
\label{fig:overview}
\end{figure*}
\paragraph{USB Attacks}
Exploits on the USB protocol and implementations of it (e.g., on
hosts, peripherals, and controllers) may occur from the physical layer upwards.
An example of a physical layer attack could be a malicious USB device that
destroys the bus by using out-of-specification voltages and currents via
large capacitors~\cite{usb-zapper}. An example of a more subtle attack is a ``BadUSB'' attack~\cite{badusb}.
This attack acts
completely within the USB protocol and abuses the trust of users and the lack
of USB device authenticity. During the USB enumeration phase, the USB host will
query the device in order to discover its functionality via descriptors
(e.g., keyboard,
storage, webcam, etc.), but a BadUSB device will misrepresent itself as
an unlikely device type. In concrete terms, a flash drive could claim itself as
a keyboard or network device without consequence. This mismatch between
physical device and presentation of capabilities could be used to socially engineer users~\cite{tischer2016users} who
would trust a keyboard differently than a flash drive (e.g., not anticipating keystrokes from their flash drive).

What actually constitutes a malicious or misrepresenting USB device is simply a
malicious application of the USB protocol.  This application which, depending
on the device, runs on hardware, a microcontroller, a CPU, or any
combination of these determines how a device functions when exposed to the USB
protocol. \papername focuses specifically on the software that runs on USB
microcontrollers, in particular microcontrollers that utilize the \emph{8051
architecture}.

\subsection{Firmware Analysis}
Microcontroller devices are often overlooked, but with the explosion of embedded and IoT 
devices, these are becoming woven in to the fabric of our modern
day society. It is thus vital to have methods for demonstrating their
trustworthiness. USB devices represent one of many classes of devices
that run firmware, but are particularly interesting to study, both due to
the widespread deployment of existing devices, and because in newer
computers, many interfaces are being replaced with USB-C connections to
provide all peripheral functionality. While the physical signal
characteristics may differ between USB connection types and protocol
versions, the same security issues (e.g., the USB-IF states that
users are responsible the security of USB devices) remain present in all
devices.

\paragraph{8051 Architecture}
The Intel MCS-51, also known as the 8051, was developed by Intel
Corporation in 1980~\cite{wharton1980introduction} for use in embedded
systems. Despite the 8051 being nearly 40 years old, it remains a popular
design due to its reliability, simplicity, and low cost, and can be
purchased in lightweight microcontrollers or embedded into FPGA/ASIC
designs via an IP core.  The 8051 is an 8-bit microcontroller based on a
Harvard architecture and contains four major memory spaces: code, on-board
memory (RAM), external RAM (XRAM), and Special Function Registers (SFRs).
The 8051 contains 128 bytes of RAM and its extended variant, the 8052,
contains 256 bytes and additional SFRs and timers. The 8052 has no
instruction set differences from the 8051. This microcontroller has 32
registers spread across four memory-mapped banks and many SFRs for
controlling the processor's peripherals. The notable SFRs are \texttt{PSW},
which controls the register banks and contains the carry flag, and the
\texttt{IE} register, which controls interrupt functionality.


\paragraph{Intermediate Representation}
In order to analyze firmware, machine code needs to be translated, or
\emph{lifted}, into an architecture-independent representation.  An
Intermediate Representation (IR) aims to be semantically equivalent to the
underlying machine code, while being generic enough to support many
different instruction operations across many architectures.  There are many
existing IRs in use today, with each having a specific focus and purpose.
Therefore, an important design choice in binary analysis is the IR to use.
By supporting a single unified IR, the footprint of the supporting analysis
is smaller and possibly simpler. The alternative would be to have an
architecture-specific engine, but this approach would require a rewrite of
the engine and all of the analyses built upon it when targeting a new
architecture.

Two notable intermediate representations are LLVM and VEX IR. The former is
used by the LLVM compiler framework while the latter is used by the
Valgrind dynamic analysis framework. A major difference between the IRs is
that LLVM is meant for \emph{compilation} (top-down) while VEX lifts
machine code (bottom-up) and then drops back down after instrumentation.
Both IRs support a variety of architectures and are supported by symbolic
execution engines (\fie and \angr respectively). However, to 
our knowledge, prior to this work neither LLVM nor VEX had any support for
the 8051 ISA.

\paragraph{Symbolic Execution}
Symbolic execution \cite{King76} is a program analysis technique that
represents input values (e.g., registers or memory addresses) as variables
that may hold \emph{any} value. As a program is executed, symbolic values
are propagated as a side effect of updates.  Symbolic constraints
encountered in conditional branch instructions are accumulated in what is
called a {\em path condition}.  When a conditional branch instruction
is evaluated, the decision whether to take a specific branch is determined
by the
satisfiability of the path condition in conjunction with the
symbolic constraints of the branch condition. For each feasible branch, a
clone of the current execution state is created and the path condition is
updated with the symbolic branch condition. 

Symbolic execution engines suffer from the {\em path explosion} problem as
the number of paths to be considered is exponential in the number of branches
considered. Therefore, state-of-the art symbolic execution engines come with a
variety of path exploration strategies such as random selection and
coverage-based progress. Although symbolic execution has emerged as an effective
white-box testing technique, we use it to determine reachability
of states that can help us understand various characteristics of a USB device.

\section{Overview of \papernamen}
\label{sec:overview}


\papername is an extensible framework
for execution and semantic analysis of firmware images. 
The primary purpose of \papername is to act as a \emph{semantic query engine}
via a combination of static and symbolic analysis.
Unlike other solutions that rely on 
a device's actions~\cite{angel2015defending} or on human
interaction~\cite{tbb15} to determine its
trustworthiness. \papername examines the device firmware 
to determine its capability for generating potentially malicious behavior.
In general, determining if a device is malicious or benign via its firmware is
a difficult task because of the many different device architectures and operating environments.
As a result, we have specialized this tool to aid in the analysis of binary USB controller firmware.

\papername synthesizes multiple techniques to effectively reason about USB firmware.
Its most significant component is a \emph{symbolic execution engine} that
allows binary firmware to be executed beyond simple concrete inputs.
Another involves static analysis on assembly and IR instructions.
The glue that binds these components is domain knowledge.
Through informing \papername about specific protocols such as USB,
we are able to relate and guide the execution of the firmware binary
to what we publicly know about the protocol.
This allows analysis to begin from commonly available generic data
-- in our case, USB descriptors. From there we
can begin to unravel more about the firmware's specifics, such as whether
this data is actually referenced during operation.

\paragraph{High-Level Flow}
\autoref{fig:overview} illustrates \papername's process of collecting
information, analyzing it, and characterizing the potential malice of a device. Normally when a USB device gets plugged in, the operating system will enumerate the device and, based on the class, interface it with the appropriate subsystems.
Instead of sandboxing or requesting user input in
order to determine how to classify a device, \papername directly examines the
device firmware in order to query this information directly. \papername
begins its analysis by performing an initial pass for {\em signatures}
relating to USB operation, such as interfaces \circlenum{1}.
The type of
interfaces that are expected to be supported by devices of the claimed identity
are passed to the static analysis stage, which identifies memory addresses and
instructions that would be relevant to an attack scenario. The static analysis
component supports a variety of domain specific queries that can be used for (1)
determining whether a device is malicious and (2) providing semantic slicing of the
firmware for facilitating more accurate analysis such as symbolic execution.
Memory references to these descriptors are discovered and any valid code
location that is found is marked as a ``target'' for the symbolic execution
stage.  Upon finding these descriptors, the reported product, vendor IDs,
configuration, and interface information are parsed based on {\em USBDB}, a
database of operational information that we have extracted from the Linux
kernel. Such parsing allows device firmware to be
correlated against expected USB behavior.\circlenum{3}

The next stage is symbolic execution, which can provide a more precise
answer on reachability of instructions of interest or the {\em target
instructions} that have been computed by the static analysis stage based on
the specific semantic query \circlenum{2}. \papername is able to search for
any instance of USB misbehavior or non-adherence to the protocol, given the
appropriate queries. As a demonstration, we currently support two types of queries focusing on the BadUSB attack.
The first type of query is about potential interfaces the device may use
during its operation, e.g., ``Will the device ever claim to be an HID
device?'' The second type of query relates to consistency of device
behavior based on the interface it claims to have, e.g., ``Will the device send data entered by the user or will it use crafted data?''.
The first query consists of a target reachability pass that attempts to
reach the code referencing
USB descriptors. When these locations are reached, the found path conditions
will demonstrate the key memory addresses and values required to reach these
locations, implying the ability to reach this code location during runtime on a real device. The
path conditions required to reach this location in the code further inform
\papername about the addresses being used for USB specific comparisons.
For example, if an HID descriptor is reached, then we should expect to see a
memory constraint of \texttt{MEM[X] == 33}. Additionally, if an expected mass
storage device firmware reaches an HID descriptor, this could be an indicator of
malice or other anomalous behavior.
The second query is a check
for consistency regarding USB endpoints. For example, if an endpoint for
keyboard data flow is observed to reference a concrete value, this could indicate static
keystroke injection.
These gathered facts about the binary are used to construct a model of
operation \circlenum{4} that is compared against an expected model of
behavior \circlenum{5}.
This model is built from the known device VID, PID, and interface descriptors that are extracted from the binary and searched in the USBDB.
Finally the results are reported for further review \circlenum{6}.

\paragraph{Core Components}
In lieu of writing a symbolic execution engine from scratch, we used the well-established
engines developed by the \fie~\cite{DMR13} and \angr~\cite{SWS16} projects. In order to
target these engines towards USB firmware, we first developed the underlying
architecture support for each engine. This consists of machine definitions
(registers, memory regions, I/O, etc.) and an 8051 machine code to IR translator
known as a \emph{lifter}. We opted to use two
different backends to better understand the strengths of each approach. These are
detailed further in \autoref{sec:design}.  \angr utilizes VEX IR,
which was originally created for Valgrind~\cite{vex-ir}~-- a dynamic program instrumentation
tool. \fie embeds the \klee symbolic execution engine~\cite{CDE08}, which uses LLVM IR,
originally developed by the LLVM project as a compilation target.
The IR syntax of VEX and LLVM differ greatly, but the underlying semantics of both 8051 lifters are virtually equivalent, with some exceptions.\footnote{Each IR has different operations available to it. VEX IR has many specific operations relating to vector instructions and saturating arithmetic, while LLVM has no saturating IR operations to speak of. The specificity of the underlying IR can affect analysis tool understanding of program itself.} The complexity of these binary lifters is extremely high as they must map each and every subtle architectural detail from a reference manual written informally to the target IR and architecture specification. Beyond the likelihood of omitting a critical detail, some instructions may not easily map to a target IR, causing many IR instructions to be emitted. This is a major factor in the speed of the underlying execution engine and having an optimized and well-mapped IR can improve performance.

\paragraph{Threat Model}
In designing a firmware analysis tool, we must make assumptions about the
firmware images being analyzed. \papername assumes that images
being analyzed are \emph{genuine}, meaning that they have not been
specifically tampered with in order to interfere with analysis
during firmware extraction or the build step. Additionally, \papername
does not support obfuscated firmware images with the purpose to hide
control flow or memory accesses. We otherwise assume that the adversary has
the ability to arbitrarily tamper with the firmware prior to its placement
on the device or at any time prior to running \papername.
During analysis,
\papername does not consider attacks on the USB protocol, vulnerabilities in
the host drivers, or the physical layer (e.g. USB-C combined with Thunderbolt
to perform DMA attacks) as protocol analysis and driver protection are
handled by other solutions.  
We assume that the adversarial device can operate within the USB
specification, but can 
potentially masquerading as one or more devices.
In summary, \papername assumes firmware is genuine, unobfuscated, and
non-adversarial to the analysis engine. We discuss future potential
additions to the framework to further strengthen the adversarial model
in~\autoref{sec:disc}.

\section{Design and Implementation}
\label{sec:design}

\papername leverages existing symbolic execution frameworks, which allows us
to focus on identifying malicious USB firmware.\footnote{There were some
circumstances where additional efforts were required with the frameworks;
these issues are discussed in \autoref{sec:disc}.} 
The primary new components we developed to support this analysis
consist of two 8051 lifters to IR, modifications to \angr to support
interrupt scheduling, and the development of semantic firmware queries with
a basis in the USB protocol.

\subsection{8051 Lifting to IR}
\begin{figure}[t]
\centering
\includegraphics[width=\linewidth]{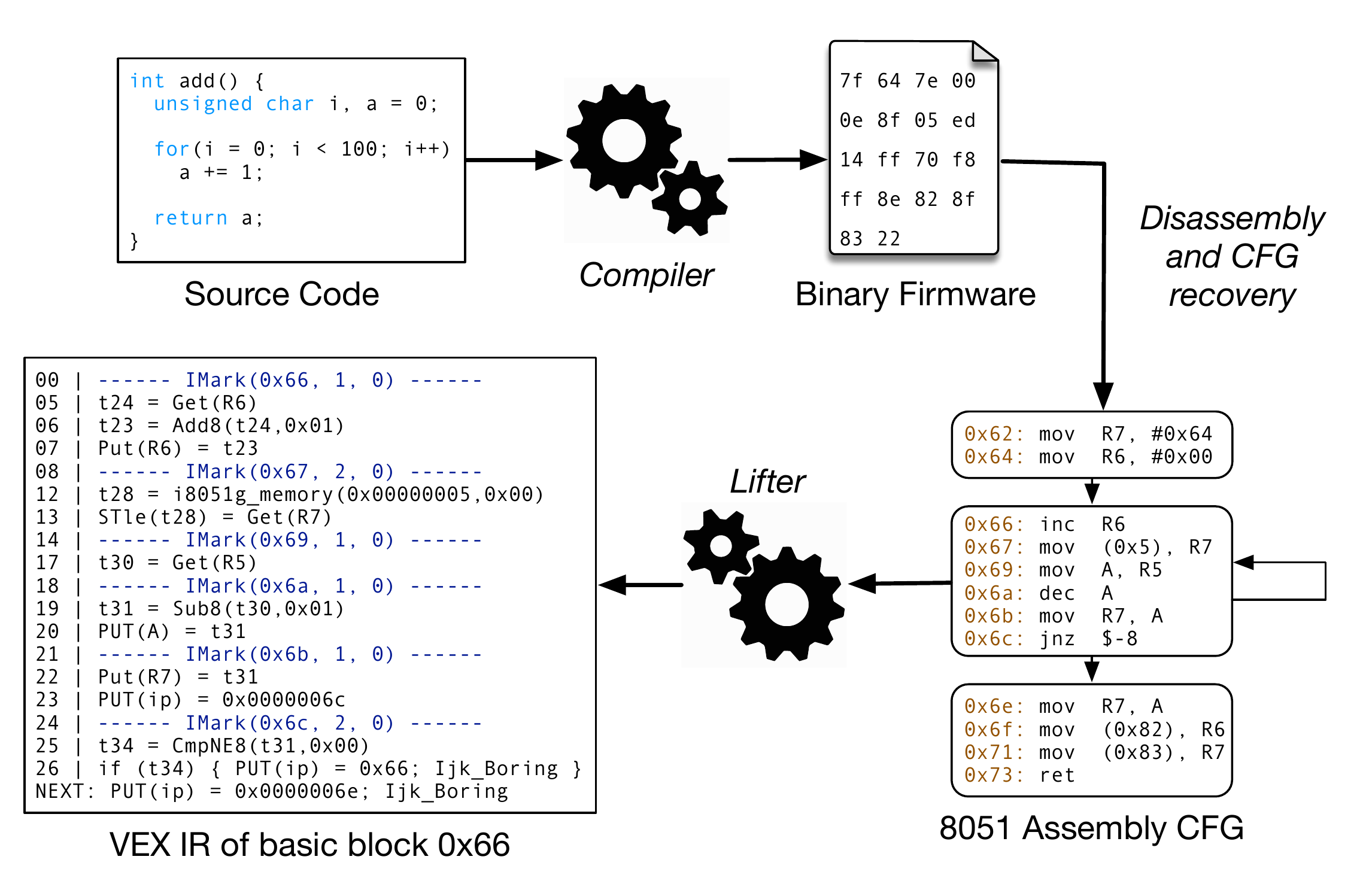}
\caption{The relationship between source code, 8051 binary firmware, and lifted IR with VEX as the example.}
\label{fig:binarytoir}
\end{figure}

In order to reason about firmware, it is necessary to represent it in a format that is amenable to symbolic analysis. The process of converting binary firmware into a corresponding IR is shown in~\autoref{fig:binarytoir}. To facilitate this process, we built two lifters for 8051 binaries: a lifter to VEX IR for use with \angr and one for LLVM IR for use with \fie. Both lifters were written to lift 8051 machine code to the equivalent or nearest representation in their respective IRs. Writing lifters is non-trivial because of the substantial number of instructions involved and the precision required. Every instruction and sub-opcode needs to be accurately mapped to the target IR.

The 8051 has 44 different mnemonics (e.g. \texttt{ADD}, \texttt{INC}, \texttt{MOV}, \texttt{LJMP}) across 256 different opcodes (e.g. \texttt{INC A} versus \texttt{INC R0}), each of which may define a one-byte, two-byte or three-byte instruction. For each opcode, the decoding pattern with operand types were manually encoded into a 256 entry table. Some example operand types included the accumulator (A), an 8 or 16-bit immediate, an address, or a general purpose register. Even with an architecture significantly less instruction-rich than Intel's current x86 and x86\_64 architectures, this lifting process took months.

Any inaccuracy in lifting, no matter how subtle, may cause code paths to be ignored or incorrect values to be calculated during symbolic execution. Processor documentation is written to be read by humans, not machines, meaning that it is easy to miss technicalities while transcribing the instructions. For example, while lifting a specific 8051 \texttt{MOV} instruction, we later noticed that unlike all other instructions, which followed the form of \texttt{[op, dst, src]}, it is the the \emph{only} instruction to have the form of \texttt{[op, src, dst]} in the binary instruction stream. This detail was missed on the first lifting pass and caused subtle incorrect execution, leading to time-consuming debugging.
Ideally, processor documentation would also be accompanied by an \emph{instruction specification}. Such a formal definition of instructions, which would include their encoding and functionality, could possibly lead to an \emph{automatic lifter generator} to be created.

There are very few disassemblers available for the 8051 architecture. We  used D52\footnote{Available from http://www.bipom.com/dis51.php} and mcs51-disasm\footnote{Available from https://github.com/pfalcon/sdcc/blob/master/support/scripts/mcs51-disasm.pl} disassemblers, in addition to our own new, table-based 8051 disassembler built into our VEX lifter and exposed to \angr via a Python wrapper we wrote called \texttt{py8051}.
To support our symbolic execution engines, the disassembled instructions are mapped to their corresponding IR.  This mapping allows the engine to emulate hardware while revealing it in detail to the analysis engine. At this stage, additional information regarding the instruction set architecture and memory layout of the device is added. On 8051, a distinction between memory regions is required as there are numerous types of memory accesses, including memory-mapped registers, external RAM, and code bytes.

\paragraph{\fie Backend.} 
To facilitate memory analysis, we built a translator that remaps 8051 machine code to LLVM IR for further use within \fie. The translator has two main components -- Dictionary and Memory Layout. The dictionary maps 8051 instructions into their corresponding LLVM IR sequence, e.g., for an \texttt{add} instruction, the IR mapping is to \texttt{LOAD} the operands, \texttt{ADD} the loaded values, and \texttt{STORE} to the destination operand location. Memory Layout contains addresses of registers and special memory locations. This memory layout is loaded by \fie in order to correctly handle symbolic memory and normal memory operations. \fie also requires interrupt subroutines to be specified. This requires an extra step to specify function boundaries by matching them against the compiled and assembled file.

\fie was built as a source level analysis tool, operating on LLVM bytecode as generated by the {\tt Clang} compiler. As a result, it was not designed to load raw binaries. Instructions that refer to code bytes in binary firmware may not be used properly without a direct reference to raw code bytes. These bytes are referred to for accessing constant data (e.g. USB descriptors and jump tables). Since \fie does not load code bytes, it does not support these instructions. To address this issue, we modified \fie to load binary firmware to handle access to code regions. This allowed us to properly symbolically execute the destination of the jumps, therefore increasing our code coverage.

In total our 8051 to LLVM IR lifter consisted of 1,967 lines of Java with 803 of 8051-to-IR specification mappings. Our direct changes to \fie consisted of 4,716 lines of C++.

\paragraph{\angr Backend}
The \angr binary analysis platform is meant to be used in projects as a Python library. It supports many binary formats (e.g. ELF, PE, MachO, raw binary) and many processor architectures (e.g. ARM, AArch64, x86, x86\_64, PPC) out-of-the-box. Despite this, during \papername's initial development, no processor architecture with a word size less than 32-bits was supported. This has recently changed with the addition of AVR support. Going forward with \papername, we opted to utilize \angr as a library and make as little modifications to the core, architecture independent code as possible. This would allows us to rebase \papername specific modifications more easily when a new version of \angr is released.

The \angr project is made up of three major subprojects -- a binary file loader
\emph{CLE Loads Everything (CLE)}, a symbolic execution engine \emph{SimuVEX},\footnote{Our version of \angr (\texttt{07bb8cbe}) is before SimuVEX and the core were merged.}
and an architecture definition repository, \emph{archinfo}. \angr composes these and many
other code bases and provides Projects, Paths, PathGroups, and many more
abstractions to aid in analyzing binaries.
In order for \angr to support the 8051 architecture, we developed a VEX IR lifter, firmware loader in CLE, architecture definition file in archinfo, disassembler wrapper (\texttt{py8051}), and IR-specific CCalls in SimuVEX for load and store addreses.\footnote{CCalls are an IR expression that acts as a callback into a function. It is primarily used by VEX to assist in supported complicated IR blocks, but we utilize it to resolve 8051 memory addresses to regions.}
In total we added 917 lines of Python code to core \angr subprojects, 623 lines of C for our 8051 disassembler, 2,850 lines of C for our VEX lifter along with 343 lines of 8051-to-IR tests. \papername's usage of \angr as a library, which included the frontend, interrupt scheduling and 8051 environment definitions amounted to 3,117 lines of Python and C.

The architecture loader consisted of a mapping between the 8051 VEX guest state structure, which contains the 8051 CPU registers, to human-readable register names. The firmware loader we added was responsible for mapping in the binary's code section and automatically discovering ISRs for later scheduling.
In order to have parity with \fie, an ExecutionTechnique was written for \angr
to support the dynamic scheduling of interrupts. The 8051 architecture uses an
interrupt vector table in which the first bytes of a firmware image are
trampolines to an Interrupt Service Routine (ISR) or a \texttt{reti}, which
means no ISR. Knowledge of the 8051 interrupt semantics were used to limit when
they and which ones were run.
To improve the execution time of firmware binaries, we created the concept of
code coverage in order to give our engine feedback on path performance.
Additionally, we created a randomized cooldown for ISR scheduling order to
facilitate binaries that have expensive ISRs. One more heuristic we incorporated
was the ability to detect looping paths with a fixed threshold. This
functionality was already built into \angr, but we utilized it to continually
prune paths not making any progress.

\subsection{VID/PID Based Inference}

To figure out what a firmware image would look like to the operating system when flashed on to a device,
we simulate how the operating system recognizes a USB device and loads the corresponding driver.
Ideally, one would expect to find all the necessary information about the device using its Vendor ID (VID) and Product ID (PID).
Unfortunately, this only works for a small portion of USB devices, an exception being USB class devices.
These devices usually follow a specific USB class specification, e.g., USB Mass Storage.
The benefit of having USB class devices is that the OS can provide a \emph{general purpose} driver to serve devices from different vendors -- as long as they follow the USB class specification.
In this case, the \texttt{bDeviceClass} solely determines the functionality of the device.
Another exception comes from USB composite devices.
These devices expose \emph{multiple} interfaces and functionalities to the operating system.
For instance, a USB headset may control an HID interface and three audio interfaces.
This is good example of where a simple VID/PID pair is just not enough to find two different drivers at the same time.

To solve these issues, we extract all the USB device matching information from the Linux 4.9 kernel, and save it as a ``USBDB''.
We have also fully implemented how the Linux kernel uses the USB device descriptor, configuration descriptors, and interface descriptors
to match a device driver.
Besides simple VID/PID matching, there are another nine matching rules~\footnote{All matching rules are listed in~\autoref{lst:usbdb} in the Appendix.} to find the corresponding drivers.
With the help of USBDB,
we may anticipate the behavior or functionality of the device firmware precisely, without having it interact with the actual OS.

\subsection{Semantic Analysis} 

In this section we explain our developed algorithms that employ a combination of static analysis and symbolic execution to  compute and check reachability of candidate target instructions for the semantic queries. Static analysis algorithms presented in this section refer to elements from the LLVM IR. Due to space restrictions, we omit their adaptation to the VEX IR.

\paragraph{Query Type 1: "The Claimed Identity?"} 

A USB device communicates with the host by responding to requests. Among those requests, \texttt{GetDescriptor} requests have a special role as it is when the device 
tells the operating system about itself. Depending on the type of the descriptor, the device would respond with specific information such as general device characteristics and configurations. For HID devices, for example,  additionally a report descriptor would be requested so that the host knows how to interpret data from the HID device. What is common among these information exchanges is that the device communicates with the host through its endpoint 0 (EP0), which corresponds to one of the ports of the device. So it is reasonable to assume that the firmware would be copying the device descriptor, the configuration descriptor, and functionality specific information, such as the HID report descriptor, to the same buffer.

\autoref{fig:algfintarget}  leverages this fact to identify candidate instructions that may be copying functionality specific information, e.g., HID report descriptor. The first step  is to identify constant parts in all these descriptor types and scan the data segment of the binary for potential memory locations that may hold these descriptors (lines 2 - 15). Then, it runs \autoref{fig:algcopyconstmem}, which is an under-approximate points-to analysis for the LLVM IR, to propagate constant memory accesses. {\tt store} instructions that copy from candidate configuration descriptors or candidate device descriptors are used to compute the set of potential memory locations that correspond to EP0 buffer (lines 17 - 25). Finally, instructions that copy data from candidate HID report descriptor buffers to the candidate EP0 buffers are identified as the target instructions (lines 26-32) and are returned as output along with the candidate EP0 addresses. 

\autoref{fig:algcopyconstmem} tracks data flow among memory locations by keeping track of the address values stored in
    or indirectly accessed via memory mapped registers. To achieve this, it associates a tuple with the source and destination of instructions, when applicable, and stores in  a map $M$ (line 4). The first component of the tuple represents a potential address value and the second component represents a tracked address value, which represents the memory location from which the data originates from. At the initialization stage, every instructions' source and destination are mapped to ($\bot,\bot)$ pairs (line 4). Then the algorithm locates store instructions that copy constant values to memory mapped registers and stores in a work list (lines 5-11). The items in the work list are processed one at a time until the work list becomes empty. 
For each instruction, it finds uses of the destination value of instruction $i$ and propagates the tuple $M(i.dst)$ based on the type of the  dependent instruction. Case 1) {\tt getelementptr} instruction (lines 27-29): Since this instruction is used to compute a pointer using an index value, the first component of the tuple $M(i.dst)$ becomes a tracked address and, hence, copied to the second component in the generated tuple\footnote{Since in our lifting of 8051 to LLVM IR  {\tt getelementptr} instructions use  0 as the base address, we do not need to perform any address computation and use the index value as the intended address.}.  The first component of the generated tuple is a $\bot$ as we do not try to keep track of values stored in locations  other than the memory mapped registers. Case 2) Other instructions\footnote{To simplify the algorithm, we did not
consider the arithmetic operations which can also help propagate constant values.} (lines 19-26): The tuple is copied as is because of the fact that the instructions {\tt store}, {\tt zext}, and {\tt load}  preserve the value as well as the tracked address. For {\tt store} instructions, the use dependence may be due to the source or the destination and therefore, we update the appropriate item whereas for all other instruction types we only propagate tuples to the destination. A dependent instruction $ui$ is added to the work list as long as it is not a {\tt store} instruction with a destination that is not a memory mapped register.
It is important to note that this is not a fix-point computation algorithm as an instruction is visited at most twice (lines 15-18) and, hence, it is an under-approximate points-to analysis.

\begin{table}
\begin{tabular}{lll}
{\bf Instruction} & {\bf Source} & {\bf Destination} \\ \hline 
L1:   mov dptr,\#X276c  & NA & (X276c,$\bot$) \\
L2:   movc a,@dptr &  ($\bot$,X276c) & ($\bot$,X276c) \\
L3:  mov r4,a & ($\bot$,X276c) & ($\bot$,X276c) \\
L4:  mov dptr,\#Xf1dc &  NA & (Xf1dc,$\bot$)\\
L5:  movx @dptr,a & ($\bot$,X276c) & ($\bot$,Xf1dc) \\
\end{tabular}
\caption{Value and tracked address propagation using \autoref{fig:algcopyconstmem} for a sample 8051 assembly code block.}
\label{table:valueprop}
\end{table}

To demonstrate the value propagation,  consider the sample 8051 code block (avoiding the rather lengthy translation to LLVM IR) given in Table \ref{table:valueprop}:
Data is moved from address {\tt X276c} to address {\tt Xf1dc} at line {\tt L5}. In instruction {\tt movx @dptr,a}, the source is a register, {\tt a}. We are interested in neither {\tt a}'s address nor its value. However, what we are interested is the address that it received its value from. Similarly, we are interested in the address that {\tt dptr} is pointing to. The indirect addressing happens at {\tt L2} and at {\tt L5}, which cause the values,  X276c and Xf1dc, to become tracked addresses, respectively. In the context of \autoref{fig:algfintarget}, {\tt L2}  may represent reading from a configuration descriptor as in line 19 or a device descriptor  as in line 22. If so, if the tracked  destination address in line {\tt L5} is a constant then it is added to the set of candidate endpoint 0 addresses as in line 20 or line 23.

\begin{algorithm}[th!]
\begin{footnotesize}
\begin{algorithmic}[1]
\State {\bf FindDevSpecInfoToEP0}($F$: Firmware, $isAReg$: Memory Mapping of Registers, $type: USB protocol$)
\State $candDD \gets \emptyset$
\State $candCD \gets \emptyset$
\State $candFuncSpec \gets \emptyset$
\For{each memory location $m \in F.AddressSpace$}
  \If{$m[0] = 0X1201$}
     \State  $candDD \gets candDD \cup \{m\}$ 
  \ElsIf{$m[0] == 0X0902$}
      \State $candCD \gets candCD \cup \{m\}$
  \Else
     \If{$((type = HID$ AND $m[0] == 0X05010906)$ OR
           ($type = MASS\_STORAGE $ AND ...) OR ...} 
       \State $candFuncSpec \gets candFuncSpec \cup \{m\} $   
     \EndIf    
  \EndIf
\EndFor
\State  $M \gets ${\bf PropConstMemAccesses}$(F,isAReg)$
\State $EP0_1, EP0_2 \gets \emptyset$
\For{each store instruction $si \in F.Instructions$}
  \If{$M(si,src).second \in candCD$ and $M(si.dst).second \not = \bot$} 
     \State $EP0_1 \gets EP0_1 \cup  \{M(si.dst).second\}$
  \EndIf
 \If{$M(si,src).second \in candDD$ and $M(si.dst).second \not = \bot$}
   \State $EP0_2 \gets EP0_2 \cup  \{M(si.dst).second\}$
 \EndIf
\EndFor
\State $targetInsts, ep0 \gets \emptyset$
\For{each store instruction $si \in F.Instructions$}
  \If{$M(si,src).second \in candFuncSpec$ and $M(si.dst).second \in (EP0_1 \cap EP0_2)$}
   \State $targetInsts \gets  targetInsts \cup \{si\}$
   \State $ep0 \gets ep0 \cup \{M(si.dst).second\}$
  \EndIf 
\EndFor 
\State {\bf return} ($targetInsts$,$ep0$)  
\end{algorithmic}
\end{footnotesize}
\caption{An algorithm for finding candidate instructions that copies functionality/protocol specific information to the EP0 buffer.}
\label{fig:algfintarget}
\end{algorithm}

\begin{algorithm}[th!]
\begin{footnotesize}
\begin{algorithmic}[1]
\State {\bf PropConstMemAccesses}($F$: Firmware, $isAReg$: Memory Mapping of Registers) 
\State Let $isAReg: F.AddressSpace \to \{true,false\}$
\State {\em Output:} $M:  F.Instructions \times \{src,dst\}  \mapsto N \cup \{\bot\} \times N \cup \{\bot\}$
\State $M \gets \lambda i,j. (\bot,\bot)$
\State $worklist \gets \emptyset$
\For{each store instr. $si$ in $F.Instructions$}
   \If{$isAConstant(si.src)$ and $IsAReg(si.dst)$}
      \State $worklist \gets worklist \cup \{si\}$
      \State $M \gets M[(si.dst) \mapsto (Value(si.src),\bot)]$ 
   \EndIf
\EndFor
\While{$worklist$ not empty}
     \State $i \gets worklist.remove()$
     \For{each intra-procedural use $ui$ of $i$}
        \State $srcdef \gets M(ui,src).first \not = \bot or M(ui,src).second \not = \bot$
        \State $dstdef \gets M(ui,dst).first \not = \bot or M(ui,src).second \not = \bot$
       \If{($isAStore(ui)$ and $srcdef$ and $dstdef$) or
            ($!isAStore(ui)$ and $srcdef$ or $dstdef$) }
           continue
       \EndIf 
       \If{$isALoad(ui)$ or $isZext(ui)$}
                \State $M \gets M[(ui,dst) \mapsto M(i,dst)]$
       \ElsIf{$isAStore(ui)$}
           \If{$i.dst$ defines $ui.dst$}
              \State $M \gets M[(ui,dst) \mapsto M(i,dst)]$
           \Else // $i.dst$ defines $ui.src$
                  \State $M \gets M[(ui,src) \mapsto M(i,dst)]$   
           \EndIf
       \ElsIf{$isGetElementPtr(ui)$}
             \State $M \gets M[(ui,dst) \mapsto (\bot,M(i,dst).first)]$   
       \EndIf          
       \If{!$isAStore(ui)$ or $isAReg(ui.dst)$}
               \State $worklist \gets worklist \cup \{ui\}$       
       \EndIf        
      \EndFor
\EndWhile
\end{algorithmic}
\end{footnotesize}
\caption{Algorithm for propagating constant memory addresses.}
\label{fig:algcopyconstmem}
\end{algorithm}

\paragraph{Query Type 2: "Consistent Behavior?"}

A USB device that claims to have certain functionality is one thing, but whether it actually carries out that function is another. Therefore, it is important to check firmware for behavior that is consistent with the claimed functionality. As an example, a USB device that is claiming to have HID functionality and sending keys that are not actually pressed and then loaded in from a I/O port is not behaving consistently. To detect such inconsistent behavior, we need to define what would  be consistent first. Obviously, this requires considering specific functionality as, for example,  what is consistent for HID may not be consistent with a Mass Storage device.

\begin{algorithm}[th!]
\begin{footnotesize}
\begin{algorithmic}[1]
\State {\bf FindSymbolicLocations($F: Firmware$, $\tau: Max iterations$)} 
\State {\em Output:} $\mathcal{P}(MemoryLoc)$
\State $WSet : Execution State \to \mathcal{P}(MemoryLoc)$
\State $symbolicLocs: \mathcal{P}(MemoryLoc)$
\Function{\bf checkLoads}{$i$:Instr,  $s$: Execution State}
    \If{$isALoad(i)$ and $i \in f.Instructions$ and $i.src \not \in WSet(s) \cup symbolicLocs$}
       \State $symbolicLocs \gets symbolicLocs \cup \{i.src\}$
       \State terminate symbolic execution
    \EndIf
\EndFunction
\Function{\bf recordStores}{i:Instr, s: Execution State}
    \If{$isAStore(i)$ and $i \in f.Instructions$ and $i.dst \not \in WSet(s)$}
       \State $WSet \gets WSet[s \mapsto WSet(s) \cup \{i.dst\}]$
    \EndIf
\EndFunction
\State $symbolicLocs \gets \emptyset$
\For{each interrupt function $f$}
\For{$i$: 1 to $\tau$}
    \State $WSet \gets \lambda x. \emptyset$
    \State register {\bf checkLoads} and {\bf recordStores} as listeners for symbolic execution 
    \State run symbolic execution on $F$ with $f$ as the only interrupt function and with $symbolicLocs$
\EndFor 
\EndFor
\State {\bf return}    $symbolicLocs$  
\end{algorithmic}
\end{footnotesize}
\caption{An algorithm for finding memory locations that should be represented symbolically.}
\label{fig:algfindsym}
\end{algorithm}

\ignoreme{
\begin{algorithm}[th!]
\begin{footnotesize}
\begin{algorithmic}[1]
\State {\bf FindSymbolicLocations($F: Firmware$)} 
\State {\em Output:} $\mathcal{P}(MemoryLoc)$
\State Let $symbolicLocs \gets \emptyset$
\For{each interrupt function $if \in F.interrupts$}
    \State Let $ReachDef: BasicBlock \to \mathcal{P}(Instr)$    
    \State $ReachDef \gets ComputeReachingDefs(if)$
    \For{each load instruction $li \in if.instructions$}
        \If{$\neg \exists di \in ReachDef(li.basicBlock)$ s.t. $di$ defines $li.source$}
            \State $symbolicLocs \gets symbolicLocs \cup \{li.source\}$
        \EndIf
     \EndFor   
\EndFor
\State {\bf return}    $symbolicLocs$  
\end{algorithmic}
\end{footnotesize}
\caption{An algorithm for finding memory locations that should be represented symbolically.}
\label{fig:algfindsym}
\end{algorithm}
}

Since we target BadUSB attacks, we focus on defining and checking for consistent behavior of HID devices. An HID device is expected to send to the host whatever it receives from the user. If, as in the case of BadUSB, it is injecting keys that have not been pressed then it could mean it is either sending data that it reads from a  buffer stored in memory or sending some constant values. How can we differentiate between such crafted buffers and those that may hold user-provided data? The key here is the interrupt mechanism. When a user presses a key, an interrupt is generated and the firmware handles the interrupt to store the specific key(s) pressed. Memory locations that are read inside the interrupts are the source of data provided by the external environment. By marking these addresses as symbolic, we distinguish addresses that are filled by the environment (as opposed to appearing statically in the binary image) and those that are not.

\ignoreme{
\autoref{fig:algfindsym}  identifies memory locations that need to be represented symbolically. The algorithm traverses each interrupt function in the firmware and computes a standard reaching definitions analysis for each basic block, which records for the set of instructions (definitions) which have made the most recent assignment (definition) to some variable based on the control flow. 
Then it analyzes each load instruction to check if there exists a reaching definition for the source address. If not, then it must be defined by the environment and so the source address is added to the set of symbolic locations.  
}

\paragraph{Finding Symbolic Locations} \autoref{fig:algfindsym}  identifies memory locations that need to be represented symbolically. Since such locations are processed in interrupt functions, the algorithm symbolically executes the firmware for a single interrupt function at a time.\footnote{Nested interrupts are currently unsupported but otherwise the 8051 \texttt{IE} register is respected when it comes to interrupt scheduling.} As paths and the corresponding execution states get generated, locations written inside the interrupt function on the current path are stored in a map, $WSet$, by a listener, {\bf recordStores}, that is registered with the symbolic execution engine. Another listener, {\bf checkLoads}, 
detects load instructions reading from memory locations that have not been written in the same interrupt function and on the current path. The source location of such a load instruction is added to the set of  symbolic values and symbolic execution is restarted with the updated set of symbolic values. For each interrupt function, this process is repeated for a given number of iterations, $\tau$.

\paragraph{When Endpoints Can Be Predicted} Another issue is identifying the endpoint address that will be used for sending HID data. The endpoint number that will be used for the specific functionality is extracted by scanning the interface descriptors that come after the configuration descriptor. To acquire the endpoint address, we can use the endpoint buffer candidates computed by \autoref{fig:algfintarget} as each endpoint is normally allocated by having a constant amount of offset from the consecutive endpoints. This constant offset is the packet size, which can be 8, 16, 32, or 64 bytes depending on the speed of the device.

\begin{algorithm}[th!]
\begin{footnotesize}
\begin{algorithmic}[1]
\State {\bf FindUnexpectedDataFlow}($F: Firmware$, , $isAReg$: Memory Mapping of Registers, $EP0: \mathcal{P}(MemoryLoc)$, $Sym: \mathcal{P}(MemoryLoc, $maxEP: int$)$
\Function{\bf checkConcAccesses}{i:Instr, s: Execution State}
    \If{$i \in targetInstrs$ and $isAConstant(i.src)$}
          \State $FlaggedAccesses \gets FlaggedAccesses \cup \{i\}$
    \EndIf
\EndFunction
\State $OtherEPs : \mathcal{P}(MemoryLoc)$
\State $OtherEPs \gets \emptyset$
\For{each $i=8, 16, 32, 64$, $k=1: maxEP$}
   \For{each $j \in EP0$}
       $OtherEPs \gets OtherEPs \cup \{j+i*k\}$
    \EndFor   
\EndFor
\State $M:  F.Instructions \times \{src,dst\}  \mapsto N \cup \{\bot\} \times N \cup \{\bot\}$
\State  $M \gets ${\bf PropConstMemAccesses}$(F,isAReg)$
\State $targetInsts \gets \emptyset$
\For{each store instruction $si \in F.Instructions$}
  \If{$M(si.dst).second \in OtherEps$}
     $targetInsts \gets targetInsts \cup \{si\}$
  \EndIf
\EndFor 
\State $counters \gets \emptyset$ 
   \For{each add or sub instruction $ai \in F.Instructions$}
       \If{exists no use $ui$ of $ai$ as a getElementPtr s.t. $ai$'s result is used as an index}
          \If{$ai.dst$ is a direct address}
           \State $counters \gets counters \cup \{Value(ai.dst)\}$
           \EndIf
       \EndIf
   \EndFor
\State Register  {\bf checkConcAccesses} as a listener and run symbolic execution with symbolic values $Sym \cup counters$
\State {\bf return} $FlaggedAccesses$
\end{algorithmic}
\end{footnotesize}
\caption{An algorithm for detecting concrete data flows to any of the endpoint buffers.}
\label{fig:algcheckunexp}
\end{algorithm}

\autoref{fig:algcheckunexp} shows how candidate endpoint buffer addresses can be used to detect concrete value flows into a potential endpoint buffer. After computing candidate endpoint buffers  based on a given number of maximum endpoints to be considered and the constant offsets (lines 7-12), it identifies the store instructions that may be storing to an endpoint buffer (lines 13-19). It also identifies {\tt add} and {\tt subtract} instructions that may be manipulating counters. If such an instruction does not have a {\tt getElementPtr} reference, then it probably is not used as an index into an array. If such an instruction's destination address can be resolved, the respective memory location is identified as a potential counter (lines 20-27). 
Such counters are often used to delay the attack and becomes a bottleneck similar to the loops for symbolic execution engines.  
All counter locations are marked as symbolic in addition to the other variables symbolic addressed that have been passed as an input the algorithm (line 28). By registering a listener, {\bf checkConcAccesses} (lines 2-6), for the symbolic execution engine, suspicious instructions that may be reading a constant value into an endpoint buffer are detected and stored in $FlaggedAccesses$. 

\begin{algorithm}[th!]
\begin{footnotesize}
\begin{algorithmic}[1]
\State {\bf FindInconsistentDataFlow}($F: Firmware$)
\State $Sym, Conc : MemoryLoc \times ContextId \to Bool$
\State $Sym, Conc \gets \lambda x,y. false$
\State $FlaggedAccesses: \mathcal{P}(Instr)$
\State $FlaggedAccesses \gets \emptyset$
\Function{\bf recordAccesses}{i:Instr, s: Execution State}
    \If{$isAStore(i)$}
       \If{$isSymbolic(i.src)$}
          \State $Sym \gets Sym[(i.dst,i.blockID) \mapsto true]$
       \Else $Conc \gets Conc[(i.dst,iblockID) \mapsto true]$   
    \EndIf
   \EndIf 
\EndFunction
\Function{\bf onSymExTermination}{}
     \State $FlaggedAccesses \gets \{ \ i \ | \ Conc(i.dst,i.blockID) \ and $ 
     \State $\ \ \ \ \ \ \ \ \ \ \ \ \ \ \ \ \exists b. Sym(i.dst,b) \ and  \ b \not = i.blockID\}$
\EndFunction
\State Register  {\bf recordSymAccesses} and {\bf onSymExTermination} as listeners and run symbolic execution
\State {\bf return} $FlaggedAccesses$
\end{algorithmic}
\end{footnotesize}
\caption{An algorithm for detecting inconsistent data flows.}
\label{fig:algcheckcons}
\end{algorithm}

\paragraph{When Endpoints Cannot Be Predicted} There may be cases when endpoints are
   setup via the hardware logic and are not easily guessed, i.e., the constant offset hypothesis fails. In such cases malicious behavior can still be detected
   by checking for inconsistent data flow as shown by \autoref{fig:algcheckcons}. The algorithm assumes that the device sometimes acts non-maliciously,  i.e., the data sent to the host is read from a symbolic location, and sometimes act maliciously, i.e.,  the data sent to the host is read from a concrete location. To detect this, we perform a pass of the symbolic execution algorithm with two listeners (line 18). Listener {\bf recordAccesses} records whether a store into a memory location get its data from a symbolic or a  concrete source along with the block identifier as the context information (lines 6-13).  Upon termination of the symbolic execution algorithm, listener {\bf checkConcAccesses}  identifies memory locations that are known to receive symbolic values in some contexts and concrete values in 
others (lines 14-17). Instructions that write to such memory locations using concrete sources are stored  in $FlaggedAccesses$ and are returned by the algorithm.

\section{Evaluation}
\label{sec:eval}

\begin{table*}[ht]
\centering
\begin{footnotesize}
\begin{tabular}{l c c cc  cc  cc  cc }
  \multicolumn{3}{r}{}& \multicolumn{4}{c}{Time to Target (seconds)} &  \multicolumn{4}{c}{Coverage At Target (\%)} \\
  \multicolumn{3}{r}{} & \multicolumn{2}{c}{\backendangr} & \multicolumn{2}{c}{\backendfie} & \multicolumn{2}{c}{\backendangr} & \multicolumn{2}{c}{\backendfie} \\
  Firmware Name (Controller) & Symbolic & Domain Spec. & Config  & HID  & Config   & HID         & Config  & HID   & Config  & HID  \\\hline
  Phison (Phison 2251-03)    & Full      & No          & --      & --      & 384.40  & 43.49     & --      & --    & 59.60   & 46.47\\
                             & Partial   & No          & 68.91   & 68.72   & 58.54 & 21.64       & 49.53   & 48.58 & 48.61   & 41.91 \\
                             & Full      & Yes         & --      & --      & 55.77 & 7.91        & --      & --    & 44.66   & 38.87 \\
                             & Partial   & Yes         & 70.28   & 70.09   & 7.68  & 5.64        & 49.53   & 48.58 & 38.88   & 36.26 \\\hline

   EzHID (Cypress EZ-USB)    & Full      & No          & 10.76   & 24.04   & --    & --          & 25.92   & 36.47 &  --     &  -- \\
                             & Partial   & No          & 9.65    & 22.07   & 63.52 & 67.04       & 25.92   & 36.47 &  42.06  &  43.08 \\
                             & Full      & Yes         & 5.33    & 11.88   & --    & --          & 11.24   & 14.45 &  --     &  -- \\
                             & Partial   & Yes         & 5.18    & 11.13   & 9.45  & 9.87        & 11.24   & 14.45 &  37.95  &  38.71  \\

\end{tabular}
\end{footnotesize}
\caption{Time for each \papername backend to reach USB-related target instructions (Query 1) for our two firmwares. The symbolic column represents the symbolic mode used to execute the binary and the domain specific column states that USB specific conditions were applied to the execution. The coverage (lower is better) is included to show the effects of partial symbolic and domain constraining optimizations. The dashes (--) indicate that the run was unable to complete due to an error.}
\label{table:timingresults}
\end{table*}

We evaluate \papername based upon two malicious firmware images and across our separate backend engines built on \angr and \fie.
One firmware binary that we analyze is
reverse engineered C code from a Phison 2251-03 USB controller
(Phison) and the other (EzHID) implements a keyboard for the Cypress EZ-USB.
A key difference between the images is that the Phison firmware is meant to act
as a mass storage device, but contains hidden code to act as a Human Interface
Device (HID), whereas EzHID acts as a normal keyboard, but injects malicious
keystrokes at run time. Our evaluation goals are to determine what USB
configurations a firmware will act as during run time in order to compare
against an expected device model and to search for inconsistent behavior of its
claimed identity.

All evaluation is performed on a SuperMicro server with 128GiB of RAM and
dual Intel(R) Xeon(R) CPU E5-2630 v4 2.20GHz CPUs for a total of 20 cores.
The \backendangr used Python 2.7.10 running on PyPy 5.3.1\footnote{In practice, we received roughly a 2x speedup over the standard CPython interpreter, at the expense of greatly increased memory usage.}
while the \backendfie used a modified version of KLEE\cite{CDE08} on LLVM-2.9.
In practice, due to implementations of the backends
themselves, \papername was only able to utilize a single core (Python 2.7 and
\klee  are single threaded).  We did not opt to orchestrate multiple processes
for increased resource utilization. Except for making the EzHID firmware malicious, we did not modify or tailor the firmware images to aid \papername during analysis.

The evaluation begins with an explanation of the firmware benchmarks we used, followed by the output of our symbolic location finder from \autoref{fig:algfindsym}, then on towards our domain informed algorithms, and finally Query 1 and Query 2 on both firmwares.

\subsection{Benchmarks}

The first firmware we used for analysis is the Phison Firmware. It was reverse engineered in to C code by \cite{badusb} and then modified to perform a BadUSB attack. The firmware initially enumerates itself as a Mass Storage device and later may re-enumerate as an Human Interface Device. After a predefined threshold count, it starts sending input data from a hardcoded script. Since, the device is now operating as a keyboard, the sent data is accepted as valid keystrokes. The Phison firmware runs on top of an 8051 core, which influenced our choice to select Intel's 8051 architecture as our initial target to analyze.

Our second USB firmware case study was based on the EzHID Sun Keyboard
firmware. In normal usage this firmware was meant to work with an EZ-USB chip
for bridging the legacy Sun keyboard to a modern USB bus. From the stock
firmware, we modified the image with a malicious keystroke injector, similar to
that of the Phison firmware. After a set delay, the firmware will begin to
inject a series of static scan codes on to the active USB bus. This interrupts
the normal flow of keystrokes from the Sun keyboard until the injection has
completed.  EzHID's firmware was chosen as it was readily available online\footnote{Available from http://ezhid.sourceforge.net/sunkbd.html}
and also compatible with the 8051 architecture (with 8052 SFR and RAM extensions).

\subsection{Symbolic Values}

One of our main contributions in this paper is the Algorithm \ref{fig:algfindsym} which finds the memory locations that need to be symbolic in order to analyze the firmware. \papername utilizes two symbolic execution engines both of which require specified symbolic memory regions. Large portions of both benchmarks are only conditionally accessible. Without the correct regions being symbolic the code cannot be properly covered and the analysis becomes incomplete. When no memory region is set symbolic the coverage achieved for Phison is 17.20\% and for EzHID it is 22.49\%. In this case interrupts are still fired but due to unmet conditions, not much of the code is executed until the code finally ends up executing an infinite loop. Since the target instructions are also guarded by conditions, the symbolic execution never reaches them. As a result, the malicious property of the firmware cannot be determined without more symbolic memory. To improve this, we use Algorithm \ref{fig:algfindsym} to set memory regions as symbolic, causing us to reach the targets.

One interesting aspect here is the contrast between our two benchmarks. Phison uses direct addressing for most of the memory reads and the conditional variables on the target path. On the other hand, EzHID uses indirect reads from memory for conditional variables in the path to target. By recording loads and stores for each path we were able to record the destination of indirect memory accesses. Our algorithm found that only 26 bytes for Phison and 18 bytes for EzHID should be set symbolic. It took one iteration for each byte of the symbolic set to get all the symbolic memory locations needed to reach Query 1 target. The minimum and maximum time taken by one iteration is respectively 3.39 and 8.42 seconds. Setting memory partially symbolic based on our algorithm increased efficiency greatly. It allowed fewer paths to be created compared to setting the full memory region symbolic. From table \ref{table:timingresults} it can be seen that we have achieved a maximum of 2x speed up in reaching targets. The algorithm helped in reducing the number of paths to execute when compared to a fully symbolic memory execution. From our tests we have seen a 72.84\% reduction in number of paths created to reach targets for Phison. A certain amount of instructions must be covered to reach the target, that is why instruction coverage does not reduce as significantly as the number of paths. But this path reduction entails less branches being created which in turn increases speed.

\subsection{Domain Informed Analysis}

\paragraph{Target Finding}
A preliminary step, before symbolically executing our firmware images, is to utilize knowledge of the USB constants and byte patterns to identify targets in the binary to execute towards. Using \autoref{fig:algfintarget}, we scan the binary image
for static USB descriptors and search for all cross-references (XREFs) to these descriptors
via load instructions. The \backendfie utilizes signature scanning and a pass
over the LLVM IR while \backendangr uses signature scanning and the built-in
\texttt{CFGFast} Control Flow Graph recovery routine to automatically generate XREFs.

Using our target finding, we identify USB configuration descriptors with
the pattern \texttt{[09 02 ?? ?? ?? 01 00]}, device descriptors with
\texttt{[12 01 00 ?? 00]}, and HID keyboard reports starting with the bytes
\texttt{[05 01 09 06 A1]}. The \texttt{??} in a pattern means that the byte at
that position can take on any value. This makes the signatures more robust
against changing descriptor lengths. \autoref{fig:patternxref} is an example extracted from the
Phison firmware image showing the clear reference to the descriptor via a
\texttt{mov} into DPTR (a 16-bit special pointer register) followed by a \texttt{movc} from the code section.
\papername would then zero in on the \texttt{0xbf4} address as that is what is reading from the descriptor address.

\begin{figure}
{\tt \footnotesize \begin{verbatim}
  X0bee:  mov     r7,#0           ; 0bee
  X0bf0:  mov     a,r7            ; 0bf0
          mov     dptr,#X30c3     ; 0bf1
          movc    a,@a+dptr       ; 0bf4

  X30c3:
          .db 0x05, 0x01, 0x09, 0x06, 0xA1,
              0x01, 0x05, 0x07 ...
\end{verbatim}
}
\caption{A snippet of assembly from the Phison firmware showing how XREFs are found from patterns.}
\label{fig:patternxref}
\end{figure}

During our development and research of \papername, we refined the dynamic
analysis process through limiting the set of symbolic data and further
constraining this limited set.  Using \autoref{fig:algfindsym}, we create a
subset of symbolic variables to be instantiated during the dynamic analysis.
Through limiting the amount of symbolic memory, the targets are reached
significantly faster.  Over-approximation of symbolic memory is guaranteed to
reach all program paths at the expense of a potentially intractable amount of
created states.  \autoref{table:timingresults} demonstrates the benefits of
selectively deciding symbolic variables in terms of analysis speed while
executing Query 1.

We optimize our analysis further by utilizing preconditioned
execution \cite{ACH11}, or USB specific domain constraints to selected symbolic
variables. By adding initial conditions before running a query, the query may
complete faster. It's also possible to over-constrain execution causing the
query to end early, run very slow, or never complete. In order to know
which constraint to apply and where, we first gather facts from found targets
with constraints already applied. By modifying these constraints with
respect to USB specific constants, it is possible to quickly reach USB-specific
or prevent reaching of less important code paths.

\subsection{Target Reachability (Query 1)}
Using \papername's knowledge of the USB protocol, interesting code offsets in
the firmware binary are identified. These targets are searched for
during a symbolic execution pass. If a target is found, the path information
will be saved and the wall-clock time and current code coverage will be noted. Information collected
includes the path history, which includes every basic block executed, all path
constraints, and every branch condition. Targets are the primary basis for
gathering additional information on firmware images. It links the previous
static analysis pass to a satisfiable path. This enables more advanced analysis
and inference of a firmware's intent.

\paragraph{Phison}

We start by looking for USB specific constants in Phison to reason about Query 1. What we found is shown in \autoref{tab:phisonxref}. Using static analysis on the generated IRs we found instructions that use the either one of the descriptors to load from. For each descriptor a set is kept that records the destination addresses where these descriptors get copied to. We took the intersection of these sets and found the possible set of EP0 address. In this case there was only one common element and the EP0 address was found to be 0xf1dc. Comparing with the source code we found that this was indeed the address of the FIFO in EP0. This enabled us to find the instruction where HID descriptor was being copied to EP0. We could reach the target in short time using Algorithm \ref{fig:algfindsym} to set symbolic regions for the analysis engines. The times shown in Table \ref{table:timingresults} shows the effectiveness of our algorithms in reaching Query 1 targets. When we do not apply Algorithm \ref{fig:algfindsym} the time to reach targets is highest. The combination of Algorithm \ref{fig:algfindsym} and domain specific constraining gives the best performance. When the size of symbolic memory region is reduced we automatically end up with fewer paths to go in. Since we determine the symbolic regions in a sound way we actually reach the target with lower number of paths to test. Also domain specific constraining further improves the performance. We restricted the path based on two factors -- USB speed change packets, which do not affect our query, and making sure to guide the execution to call the \texttt{GET\_DESCRIPTOR} function as the successor when the deciding switch statement comes. This pruning is sound for reachability testing because we combine domain specific knowledge. Using our optimizations, we achieved maximum of 7.7x speed up compared to the fully symbolic version's unconstrained execution for HID target.
Our \backendangr is not able to complete the Full version of Phison due to running out of memory, which appears to be because of path explosion. 

\begin{table}
  \small
  \begin{tabular}{ l l l l }
    Pattern Name    & Pattern & Code Address & XREF(s) \\ \hline
    DEVICE\_DESC    & \texttt{[12 01 00 ?? 00]}       & 0x302b & 0xb89 \\
    CONFIG\_DESC    & \texttt{[09 02 ?? ?? ?? 01 00]} & 0x303d & 0xbd5 \\
    HID\_REPORT     & \texttt{[05 01 09 06 A1]}       & 0x3084 & 0xbf1 \\
  \end{tabular}
\caption{The found patterns and XREFs from Phison.}
\label{tab:phisonxref}
\end{table}


\paragraph{EzHID}
Using our target finding, we identified a USB configuration descriptor, a device descriptor, and an HID report in EzHID. Then we utilized our static analysis to find
code address XREFs for all targets as shown in \autoref{tab:ezhidxref}. With the list of targets, we activated
\papername for both backends. The first pass identified the required path
conditions for reached targets, which allowed us to optimize additional runs by
constraining SETUP data packet addresses that satisfy the following
constraint \texttt{XRAM[0x7fe9] == 6} from \autoref{fig:ezhidconstraints}. \texttt{0x7fe9} 
corresponds to the second byte of the USB setup data packet which is the field
\texttt{bRequest}. By limiting this to \texttt{0x06}, we effectively constrain
the execution to paths that satisfy the \texttt{GET\_DESCRIPTOR} call. For EzHID, this eliminates all other request types, which speeds up time-to-target and further analysis.
In \autoref{table:timingresults} EzHID performs better when domain constraining is enabled, but with a partial symbolic set the time to target has little change. This is due to the shallow target, which does not have time to benefit from the partial set. \fie is unable to complete the Full version of EzHID due to a memory out of bounds error, which is a limitation of \klee's symbolic memory model. See the discussion in Section \ref{sec:kleevsangr} for a further explanation.

\begin{table}
  \footnotesize
  \begin{tabular}{ l l l l }
    Pattern Name    & Pattern & Code Address & XREF(s) \\ \hline
    DEVICE\_DESC    & \texttt{[12 01 00 ?? 00]}       & 0xb8a & 0x18b  \\
    CONFIG\_DESC    & \texttt{[09 02 ?? ?? ?? 01 00]} & 0xb9c & 0x1a4  \\
    HID\_REPORT     & \texttt{[05 01 09 06 A1]}       & 0xbbe & 0x250  \\
  \end{tabular}
\caption{The found patterns and XREFs from EzHID.}
\label{tab:ezhidxref}
\end{table}

\begin{figure}
{\tt \footnotesize \begin{verbatim}
BVS(XRAM[7fab][0:0]) != 0   // USBIRQ & 0x1 ?
BVS(XRAM[7fe9])      == 6   // bRequest - Descriptor
BVS(XRAM[7feb])      == 34  // wValueH - HID Report
BVS(XRAM[7fec])      == 0   // wIndexL - Keyboard Index
\end{verbatim}
}
\caption{The path constraints present at the execution step when the HID report was reached for EzHID.}
\label{fig:ezhidconstraints}
\end{figure}

\subsection{Consistent Behavior (Query 2)}
A second important query to vetting USB device firmware is detecting inconsistent use of USB endpoints. In a typical mass storage device, one would expect SCSI commands to arrive, data to be read from flash memory, and then transferred out over an I/O port to the USB bus. While analyzing firmware \papername treats memory reads from I/O regions (typically external RAM or XRAM) as symbolic. Therefore, a consistent firmware image for either mass storage or HID should read symbolic data, process it, and pass it along. An inconsistency would occur if a firmware writes symbolic \emph{and} concrete data to an output port from different code blocks.
\papername performs dynamic XRAM write tracking as specified in \autoref{fig:algcheckunexp} and \autoref{fig:algcheckcons}.

\paragraph{Phison}

We checked for concrete data flow in the firmware using Algorithm \ref{fig:algcheckunexp}. Since we set all inputs to be symbolic there should only be symbolic data flowing to endpoints except EP0 for descriptors. The concrete data flow to endpoints in this case entails stored data being propagated to the host. As the Phison firmware should work as a mass storage device firmware this behavior is inconsistent. EP0 found for Query 1 is used to calculate other endpoint addresses using constant offset. A threshold count of 8192 was there in the firmware. Due to this count the concrete data flow was getting delayed and our symbolic execution engines did not execute the malicious code region. That is why \autoref{fig:algcheckunexp} was extended to incorporate these counters that compare with the threshold. We used the algorithm to find the counters that may guard this execution. We found 14 more bytes of memory and included them to the already found symbolic memory regions. Once these additional memory regions were made symbolic we could reach the Query 2 target for Phison. We found constant data being copied to EP3. With the new set of symbolic memory, it took 928.56 seconds to reach the target with 69.98\% instruction coverage. There was one false positive due to a SCSI related constant being copied to EP1.

\paragraph{EzHID}
After finding the USB specific targets, this firmware does not appear
suspicious as it is supposed to be a keyboard firmware. In order to further vet
this image, we perform a consistency check on the USB endpoint usage.  This
query consists of running the firmware image for an extended period in order to
capture as many XRAM writes as possible. If an inconsistency is detected, the
results and offending instructions (and data) are printed for further
investigation. An example of malicious code that injects keystrokes is shown in
the Appendix as \autoref{lst:inconsistent}.  Using
\autoref{fig:algcheckcons}
to detect when an inconsistency has occurred, our
\backendangr will then print out the offending memory writes, their write
addresses, and the section of code that did the writing. There are some false
positives, but the most consistent violations
(more than one violation indicating many writes) will be ranked higher. We ran
Query 2 for 30 minutes to gather the results which are displayed in \autoref{tab:ezhidq2}.
The first row shows the discovered inconsistent writers, where one writes symbolic scancodes and another only concrete data from the firmware images.
The next two rows are false positives, which also have many different write sites, but the difference is that each write address only writes a single concrete value. The same writer does not have multiple violations (such as writing many different keystrokes).

\begin{table}
  \footnotesize
  \begin{tabular}{ l p{1.5cm} l p{2.0cm} }
    Write Address       & Writers & Symbolic Value & Concrete \\ \hline
    0x7e80 -- 0x7e87  & 0x991, 0xa7e & scancode[0-7]  & 0x0, 0xe2, 0x3b, 0x1b, 0x17, 0x08, 0x15, 0x10,  0x28 \\ \hline
    0x7fd4            & 0x199, 0x1b2, 0x22c, 0x1e9, 0x1e9, 0x25e, 0x6d7, 0x161 & SDAT2[7fea]  & 0x0, 0xb, 0x7f       \\ \hline
    0x7fd5            & 0x1a2, 0x1bb, 0x237, 0x201, 0x201, 0x267, 0x6d7, 0x161 & SDAT2[7fea]  & 0x0, 0x8a, 0x9c, 0xae, 0xe8, 0xbe \\
  \end{tabular}
\caption{The results of running Query 2 on EzHID for 30 minutes.}
\label{tab:ezhidq2}
\end{table}

\section{Discussion}
\label{sec:disc}
\ignoreme{
\outline{This sections will reflect upon \papername and discuss some of the quirks and issues we experienced while developing it. These can include fundamental issues involving binary analysis and symbolic execution or specific things to \papername.}
\subsection{LLVM IR}
\noindent \paragraph{{\bf Static Single Assignment:}} In {\tt LLVM IR} instructions are represented using {\tt Static Single Assignment (SSA)}. The key concept of {\tt SSA} form is that a variable can be assigned to only once. Once assigned It may be used any number of times. For a memory store operation, this entails using a new variable that has been assigned the new value from after the store. The translator has to take this into account. For our implementation a version table is kept for each register where the new variable name is the name of the register concatenated with the new version number. For memory operations the naming is easier. Since, the variables become independent once assigned a value loaded from the appropriate address, a counter counting over all memory operations suffices. The users of this variable are not affected.
\noindent \paragraph{\bf Scope:} One important thing to remember here is that all of these {\tt LLVM IR} variables are valid in the scope of one assembly instruction. The reason behind this narrow scope is that every operation loads a value from the operand memory address or in case of registers refers to that register by address. The translator has to take this into account and basically use a {\tt LLVM} load instruction for every source operand. One assembly instruction translates to one or more {\tt LLVM IR} instructions. In each such sequence of instructions it is imperative to read the source operand from corresponding address. Even if the last written value to a memory location is saved in a {\tt LLVM} variable it is necessary to read the value each time an assembly instruction requires. One reason for this is that, an interrupt can change the content of that source address. Also, indirect addressing or relative offset setting can change the value.

\noindent \paragraph{{\bf Assembly to IR:}} While {\tt LLVM IR} has the ability to represent most assembly operations into a sequence of {\tt IR} operations, it is designed for converting source code into {\tt IR}. The information available at the source level is not present in firmware binary in the same way e.g. function boundaries. As a result, a disassembled firmware does not explicitly contain all those information. Many things that are done in the hardware have to be explicitly mentioned in {\tt IR}. Which tend to make the {\tt IR} much longer in terms of number of instructions. For example an instruction such as {\tt mov dest, src } copies the content of address {\tt src} to address {\tt dest}. It can be a simple bus transaction between two locations in memory. The operation is also dependent on the word size of the architecture. Register contents are not affected by this instruction. So, while converting this {\tt mov} instruction into {\tt LLVM IR} there are a few things to consider. Data has to be read starting from {\tt src} of length determined by the word size. It may not be directly saved to {\tt dest}. The read value has to be explicitly written to the {\tt dest}. For instructions that affect the flags, explicitly each flag in {\tt IR} has to be set with the appropriate value. Flag setting cannot be determined by only the assembly instructions. Architecture dependent information is needed for this. For example in 8051 architecture it's  not just the arithmetic operations that affect carry flag, there are some rotate operations that can affect it too. Extensive knowledge of the instruction set architecture is required for this purpose.

\subsection{Function Boundaries}
\noindent {\bf Function Boundary:} {\tt LLVM IR} requires a label to be able to jump to a basic block. Jump instructions from assembly are handled via {\tt LLVM IR} branch instructions. The branch instruction requires a label as operand. The label provided has to be the basic-block name of the destination. On the other hand, the assembly instructions use an address as operand in jump instructions. For example the assembly code of a small function and it's corresponding {\tt LLVM IR} is following.
{\tt \small \begin{verbatim}
X0000:        // address 0x0
  ljmp X0005  // jump to address 0x0005
X0005:        // address 0x0005
  ret         // return	
\end{verbatim}
}
\noindent The preceding assembly would be translated as following
{\tt \small \begin{verbatim}
define void @function() {	// LLVM IR function definition
X0000:             // label X0000
  br label %X0005  
  // direct branch to label X0005
X0005:             // label X0005
  ret  }           // return
\end{verbatim}
}
\noindent While translating to {\tt LLVM IR} we have used the addresses of basic-blocks as the name. The addresses are unique to each block, which also resolves naming conflicts. For the return instruction the translation is a simple {\tt LLVM} return instruction {\tt ret}. Using functional boundaries makes it possible to translate the return instructions easily. Now if we did not have the function boundaries, the translation of the return instructions will be a lot different. For a return like that first the stack has to be popped to get the return address. That address has to be loaded in the program counter and then the program counter value will have to be used to find the destination block. It can be done using the indirect branch instruction in {\tt LLVM IR}. The example given earlier would then become as following -
{\tt \small
\begin{verbatim}
//No function definition
X0000:             //label X0000
  br label %X0005  
  //direct branch to label X0005
X0005:             //label X0005
  ...
//pop stack and store in PC not shown
  indirectbr i16* @PC, [ LABELS ]//return
\end{verbatim}
}
\noindent {\tt @PC} here refers to the program counter. {\tt LABELS} corresponds to a list of possible block names to return to. The list can be empty too. For function calls this list should be the set of all the addresses that calls the function. For ISRs this return should be to all possible addresses. Without this {\tt LABELS} the CFG does not have sufficient information. {\tt FIE} on the other hand requires the ISRs to be provided as functions. So, the functional boundaries are needed.

\noindent \paragraph{\bf Finding Function Boundaries:} To find function boundaries from the disassembled binary files the source assembly files are used as reference. The sequence of assembly instructions are matched from assembly file to find the corresponding functions from the disassembled target file. Since the functions were all in the same sequence in both files, they were matched using a sequential scan over the list of instructions. The addresses of each function is very significant. Function call instructions use the addresses of each function to call it. The call operation updates the program counter with the address of the functions which drives the execution to the entry block of the called function. Needless to say, passing of arguments are handled via a set of instructions prior to the call at assembly level. All the data pushing can be handled before or after calling the function. Function calls in firmware are not always handled via call instructions. In some cases just a jump instruction may be the one making a new function execute. This happens because of compiler optimization. In such cases the translation has to be a call instead of a branch instruction in {\tt LLVM IR}. So, while translating jump instructions the operands have to be taken into account. 

}

In this section, we discuss discrepancies between \fie and \angr, challenges with obfuscation, and features of an ideal framework for analyzing firmware.

\subsection{Adapting \fie}

\fie has built-in support for several MSP430 chips, which we used as a reference for adding the 8051 support. 
Basically, we specified the memory addresses for all registers and ports of 8051. \fie also expects special read and write functions for any memory that is declared as symbolic. These functions are normally generated automatically for architectures that are supported by {\tt Clang}. So we had to manually add these functions. 
8051 interrupt specification is introduced to \fie along with handler functions that first check whether the specific interrupt is enabled before scheduling the relevant ISR.
While the register information is available from the ISA documentation, the required symbolic memory regions are determined by \textsc{\papernamen}. 
Since Algorithm \ref{fig:algfindsym} finds each symbolic memory region iteratively, the corresponding read/write functions are created iteratively as well. On the other hand, to support 8051 interrupt firing the interrupt enable (IE) register in \fie execution engine is modified to select the right bit for interrupt enable, which turned out to be different in 8051 compared to the bit position in MSP430 that \fie initially supported.

\ignoreme{
For indirect jumps that involve switch tables, \fie needs to have access to the code regions of the binary firmware that store the switch tables. So to handle binary firmware with such indirect addressing, we have loaded binary firmware to \fie. An instruction demanding reading from the code bytes can now do so from the loaded binary. Note that, these bytes could not be read from the bitcode with the same address found from instruction operands. The reason behind this is that the addresses in the binary firmware and the generated bitcode are different. The translator produces a sequence of LLVM instructions from each 8051 firmware instruction. These LLVM instructions have different length, operand pattern and byte representation than their corresponding 8051 instruction. This entails, the correlation between bytes of the same address in binary firmware and generated bitcode is not readily known. This is not a concern when the bitcode is coming from source code through a compiler e.g. for a CLANG supported architectures. In case of \papernamen the input binary is generated from a compiler that supports 8051 (sdcc). In addition, \papernamen creates IR from the 8051 binary. This allows \papernamen to tackle indirection; one of the main concern of binary analysis. Code byte accesses are read-only.\linebreak
}

\subsection{\klee vs. \angr}
\label{sec:kleevsangr}
\klee and \angr are both symbolic execution engines, but their approaches come from different directions. \angr is a recent execution engine and it aims to be a more general purpose binary analysis platform. This means it offers code for control flow recovery, some abstract interpretation, and binary container (e.g. ELF, PE) parsing built-in. For recovering higher-level constructs from binary-only images, \angr offers a superior platform, despite its more recent development. \klee operates on LLVM bytecode, which until recently was only output from a compiler. Compilers do not have to worry about concepts such as type or control flow recovery, so targeting LLVM IR for a binary-only image was difficult. Binary firmware does not typically have any concept of types and its control flow may be masked by jump tables or indirect jumps.

As for both engines' memory models, \klee uses a linear model for every memory object and maps it to an STP array for efficient constraint solving. \angr, on the other hand, uses a more flexible indexed memory model in that it creates an immutable memory object according to the lower and upper bounds inferred from the constraint. The difference comes into play when a symbolic index into memory may suggest going out of bounds w.r.t. an existing memory object in \klee, which  flags it as a memory out-of-bounds error. This is indeed what happened when we were evaluating Query 1 on EzHID in full symbolic mode. 

\klee interleaves random path selection and a strategy to select states that are likely to cover new code as its search heuristics. A weight is computed for each process, and then a random process is selected according to these weights. These heuristics take into account the call stack of the process, whether or not the process has recently covered new code, and the minimum distance to an uncovered instruction. This interleaving technique shields the system from a case where one strategy would become stuck. Once a process is selected, it is run for a ``time slice," which is defined by a maximum amount of time and a maximum amount of instructions. Time-slicing processes helps make sure a process that is executing frequently with expensive instructions will not dominate execution time. \angr does not have any scheduling methods built in. It is left up to the user to decide which paths to prioritize. The individual execution paths in a program are managed by Path objects, which track the actions taken by paths, the path predicates, and other path-specific information. Groups of these paths are managed by \angr's PathGroup functionality, where an interface is provided for managing the splitting, merging, and filtering of paths during dynamic symbolic execution. Additionally, \angr does not collect \klee-like metrics such as code coverage, percent time spent in the solver, and instructions executed.

In a symbolic execution engine, constraint solving is a major part of checking the feasibility of a path, in order to generate assignments to symbolic variables and verify assertions. STP and Z3 are popular solvers that are used in symbolic execution engines. \klee uses STP, which only has support for bit vectors and arrays, and \angr uses Z3, which supports arithmetic, fixed-size bit-vectors, floating point numbers, extensional arrays, datatypes, uninterpreted functions, and quantifiers. Both \klee and \angr split constraints into independent sets to reduce the load on the solver.

Symbolic execution engines for binary code usually rely on transforming native instructions into an intermediate representation. LLVM generates the IR of the source code during the first step of compilation. \klee uses the IR that is generated by the LLVM compiler for C and C++. In contrast, \angr performs analysis on the Valgrind dynamic instrumentation framework (VEX) IR. VEX is a RISC-like language that is designed for program analysis and generates a set of instructions for expressing programs in static single assignment (SSA). By using VEX, they were able to provide analysis support for 32-bit and 64-bit versions of ARM, MIPS, PPC, and x86.
Beyond existing support, VEX is a binary-first IR, meaning it does not assume
a control flow graph or memory layout. With its basic block abstraction,
executing VEX IR does not require an entire binary program to be lifted
beforehand. For larger binaries, this demand-based lifting is superior in
performance and does not require any human intervention. Overall, VEX was built for binary-only targets and we believe it is the better choice for supporting new architectures.

We have explored using static analysis tools SVF \cite{SX16} and DG \cite{dg} for target identification. Although both SVF and DG provides efficient inter-procedural analysis,\footnote{LLVM's optimizer tool \texttt{opt} is mostly intra-procedural.} their analysis results were not precise enough for target finding. Specifically, the computed alias sets were very big and even {\tt store} instructions that wrote to output ports and, hence, not read in any other part of the code have been reported to have data dependencies. A closer analysis of the latter problem revealed that when compiling the Phison BadUSB firmware, the SDCC compiler used a data memory address {\tt 20h} like a register to store various flags that would affect the outcome of various branches scattered around the various parts of the firmware and appeared in five functions. Some of the accesses to this memory location were for individual bits and in others to the whole byte. Our lifter for LLVM translates accesses to individual bits by first loading the whole byte, manipulating the individual bit, and storing the whole byte back. Because of addressing in LLVM requires defining a pointer to a base region and then referring to the individual bit, translating these accesses as bit accesses would not significantly improve the precision as the pointer to the base memory region would still contribute to the imprecision in points-to analysis and, hence, to dependence analysis.

In summary, \angr shines in the analysis of pure binaries as it never assumed the availability of source or symbol files. \klee still bests \angr in raw performance (C++ vs. Python) and with a proven and well tested process execution engine, it offers a more traditional symbolic execution experience immediately without any code changes. A way to close this gap would be for the \angr project to improve and push forward a more user-friendly frontend tool to match that of \klee's. Overall, when it comes to writing code to support a new embedded architecture, \angr is the better choice due to its well-engineered modularity, active community, large amount of documentation and examples, binary-first approach, and easy and quick development cycle (no compilation times or hard crashes).
These benefits allow a researcher to quickly make progress in supporting a completely new architecture and to explore areas untouched by symbolic execution.

\subsection{Firmware Obfuscation} 
Obfuscation of code is a long standing practice to dissuade reverse engineering and to slow down attackers. With symbolic execution, obfuscated code may have an effect on the execution accuracy and performance. Unless these engines are specifically crafted to expect and handle obfuscated code, they may not be able to stand up to these code changes \cite{yadegari_symbolic_2015, banescu_code_2016}. Currently \papername's underlying symbolic execution engines \klee and \angr are not specifically designed with obfuscated code in mind. This limitation would require more engineering and testing in order to ensure reasonable performance and results in the presence of adversarial firmware.

Ignoring the timer-based delay of keyboard injection, which causes many states and delays code coverage, the two firmware images we analyzed can be considered unobfuscated. One of the effective ways to obfuscate for binary analysis would be to use as much indirection as possible for memory accesses. This would break static cross-references and prevent data flow tracking from USB constants. As an example, consider \autoref{fig:algfintarget} for identifying target instructions to find the claimed identity. If the base address of a configuration descriptor is stored to a memory location and the descriptor is always accessed by first loading from that memory location, the analysis would not be able to find any targets. A similar effect could be achieved by storing the content of the descriptors dynamically instead of being fixed at runtime. One can always use over-approximate (conservative) static analysis to overcome such obfuscation scenarios. However, for static analysis to be effective one needs strike a balance between precision and efficiency. We anticipate that domain knowledge, (e.g., the analyzed protocol(s) and the specific microcontroller architecture used), will be helpful in tuning such analyses.

\subsection{Ideal Framework \& \papernamen Limitations}
\ignoreme{
\outline{This section will discuss the major takeaways from our work. We need to generalize what we have done with \papername to other domains (more embedded systems) and discuss what the issues are that are preventing us from getting there. We want to discuss what we believe to be an ``ideal'' framework that will improve and speed up analysis on the multitude of embedded system firmwares. Our lead in will be the many challenges we experienced while developing \papername (on both the \fie and \angr sides).}
\label{sec:towards}
}
Currently \papername does not handle automatic
extraction of firmware images from the devices themselves, as this may not be
possible or vendor specific. As such, firmware images are processed offline
from public resources or extracted from a controller manually by a human. If
\papername performed automatic extraction, it would have to trust the
underlying device and USB bus to provide valid and untampered firmware images.
Even if a trusted USB bus is assumed, analysis of the firmware itself may
still be hampered by knowledgeable actors who develop adversarial firmware. For
example, if an attacker knows that \papername is being used to analyze the
firmware, he can obfuscate or cause the firmware to exhaust the resources of
the analysis engine via state explosion or delay loops. \papername will make an
effort to continue in the presence of many states by using path heuristics, but
these heuristics are fundamentally unsound. Additionally, while it is not
possible to execute data as code on the 8051 architecture -- as it is a Harvard
architecture -- it is still possible to realize weird
machines~\cite{bratus2011exploit} via Return Oriented Programming (ROP) or
Virtual Machines (VMs) via existing instructions operating on data. Any
vulnerability in the firmware that could lead to arbitrary read/write or
control flow hijacking could be abused through
self-exploitation~\cite{game-selfexploit, aol-selfexploit} in order to perform
computations not visible in the static machine code or even during run-time.

Beyond limitations of the USB protocol, it would be very useful to have an IR that reveals architectural elements, such as memory mapped or special function registers, to facilitate analysis of diverse microcontroller firmware. VEX and LLVM were not good fits for the 8051's overlapping memory model.\footnote{Code, internal RAM, and external RAM all start at address zero.} To enable precise analysis, bit level operations should also be straightforward to express in the IR, i.e.,  without resorting to tricks such as accessing the containing byte and manipulating it to achieve the intended effect

Symbolic execution will remain a key analysis component for firmware analysis. However, since most symbolic execution engines have been initially designed for analyzing user space binaries, extensions for embedded systems such as interrupts and the memory layouts have been implemented as addons to \fie and to \angr. A symbolic execution engine that provides a flexible interface for both specifying and controlling architectural aspects will be easier to engineer using domain knowledge. Also, analyzing malicious firmware would likely involve resolving intended as well as accidental\footnote{Even certain code patterns the compilers generate for efficiency may become a bottleneck from static analysis perspective.} obfuscation. Hence, symbolic execution should have flexible interfacing with other static as well as dynamic analysis components and enable reuse of facts it gathers at various phases of the analysis.

\section{Related Work}
\label{sec:relwork}
\paragraph{USB Security}
Modern operating systems fundamentally trust plugged in USB devices, as
    security decisions are left to users. As a result, operating systems and users are open to a wide variety of attacks
including malware and data exfiltration on removable storage~\cite{fmc11,sg2010,w2012},
tampered device firmware~\cite{badusb,brocker2014iseeyou},
and unauthorized devices~\cite{turnipschool}.
Solutions for applying access control to USB storage
devices~\cite{dpf2014,phs+2010,ts2010,yfq+2015} cannot assure that USB write requests are prevented from reaching the
device.  Further, protections against unauthorized or malicious device
interfaces~\cite{tbb15,sss2014} and disabling device drivers are coarse and
cannot distinguish between desired and undesired usage of a particular
interface.
Researchers have turned to virtualization as another means of providing security within USB.
GoodUSB~\cite{tbb15} leverages a QEMU-KVM as a honeypot to analyze malicious USB devices,
while
Cinch~\cite{angel2015defending} separates the trusted USB host controller and untrusted USB devices
into two domains where a gateway applies administration supplied policies to USB packets.
USBFILTER \cite{tsb+16} acts as firewall for USB and enables system administrators to only allow certain types of USB traffic.
USB devices themselves can also provide protection from malicious hosts.
USB fingerprinting~\cite{blp+14} establishes the host machine identity
using USB devices, while device-based mechanisms can attest
integrity~\cite{bmm10}, provide malware forensics~\cite{tbbr16}, provide
policy~\cite{vahldiek2015guardat,ws2010} or allow for protocol
fuzzing~\cite{Bratus2011-lv}.

\paragraph{Firmware Analysis}
\fie \cite{DMR13} is an  embedded firmware analysis platform targeting
MSP430 micro-controllers, as described previously. It leverages 
Clang's support for MSP430, requiring  
memory layout and the interrupt functions. \fie models the reactive nature
of firmware via scheduling interrupt functions at various granularities. 
AVATAR \cite{ZBF14} uses the \ste symbolic execution engine to run firmware
binaries in an emulator while forwarding I/O requests to the physical
device and processing the responses from the device through state
migration. \ste is further used~\cite{BLR15} to generate test-cases for System Management Mode interrupt handlers in BIOS implementations for Intel-based platforms. 
Firmalice \cite{SWH15} utilizes symbolic execution and program slicing to discover backdoors and their triggers in firmware images.
\cite{CZF14} and FIRMADYNE \cite{chen2016-wu} present light-weight static and dynamic analysis, respectively, for a large set of firmware collected through web-crawling.  
By contrast, \papername leverages domain knowledge of USB and embedded systems to detect BadUSB type attacks by discovering hidden functionality and inconsistent functioning.

\paragraph{Symbolic execution}
Symbolic execution engines that have been designed for analyzing systems code include EXE \cite{CGP06}, KLEE \cite{CDE08}, SAGE \cite{GLM08}, CREST \cite{BS08}, BitBlaze \cite{SBY08}, S2E \cite{CKC11}, Cloud9 \cite{CZB10}, DDT \cite{KCC10}, McVETO \cite{TLL10}, and \angr \cite{SWS16}. 
In the context of security, symbolic execution \cite{King76} has been used for
a wide number of applications; a sample of these include generating exploits for control-flow hijacking \cite{ACH11}, buffer overflows
\cite{PKF15} and other memory corruption vulnerabilities \cite{SWS16,SGS16} 
detecting exploitable bugs in binaries \cite{CAR12} and web applications
\cite{NJ14, NJ16,BHS14,CF10} proving confidentiality and integrity properties
\cite{CM12}, analyzing BIOS firmware for security \cite{BLR15}, analysis of
embedded systems' firmware binaries \cite{ZBF14,SWH15} and code \cite{DMR13},
input generation for obfuscated code through bit-level taint tracking and
architecture aware constraint generation \cite{YD15}, finding trojan message
vulnerabilities in client-server systems \cite{BCG14}, and many others. 
Our domain specific path constraining is similar to preconditioned symbolic execution in \cite{ACH11}, which constrains input length and input prefix. 
\cite{YD15} identifies inputs as those written by a library routine and then immediately read in the program, which applies to user space programs. 
Our symbolic value finding algorithm, on the other hand, is customized for interrupt driven firmware.

\paragraph{Architecture lifters}
To analyze binaries, a first step is to lift the ISA of the target device
into an IR for analysis. radare2 \cite{radare} is a reverse engineering tool
that lifts ISAs for many architectures, including 8051, into its intermediate representation, ESIL. 
However, ESIL is not popular beyond radare and its main focus is emulation to assist static analysis -- not symbolic execution.
REV.NG \cite{DPA17} leverages QEMU's TCG to lift various ISAs 
to LLVM IR. Similarly, \angr \cite{SWS16} leverages libVEX, the IR lifter of Valgrind. 
Neither QEMU nor libVEX currently support 8051. Binary Ninja~\cite{binja} is a reverse engineering platform that implements a custom IR called Low-level Intermediate Language (LLIL). A community plugin~\cite{binja-8051} lifts 8051 to LLIL, but there are no currently available symbolic execution engines for LLIL.
Precision of binary analysis depends on the accuracy of binary disassembly,
which is undecidable in the general case. It can, however, benefit from static analysis, especially    
when assumptions on compilers' behavior may not apply~\cite{DPA17}.
Lim et al. present a specification language, TSL~\cite{LR13}, for defining concrete
operational semantics of machine-code instruction sets, enabling automated generation of 
different abstract interpreters for an instruction set and retargeting to new instruction sets. 
Neither LLVM nor VEX has been target of such parameterized analysis effort. However, LLVM has received more attention from the program analysis community 
\cite{GKK15,BNS14,HMM12}.

\section{Conclusion \& Future Work}
\label{sec:conc}
\ignoreme{
\outline{This section will tie together and summarize our contributions mentioned in the abstract and intro. We are closing the paper while reflecting upon what we have learned and what other areas this paper would like to explore. Depending on space, we'd like to mention what we have in mind for extending \papername.}
\grant{Talk about automating lifters as future work}
}

This paper presented a domain informed firmware analysis that is effective in detecting BadUSB type attacks. We have formulated two semantic queries that would help reveal characteristics of a USB device. Our experiments confirm effectiveness of using domain specific heuristics for finding symbolic values and for reducing the amount of explored paths. Our lifting of 8051 instruction set to two popular IRs enabled us to leverage a source-level firmware analysis tool, \fie, and a binary analysis tool for user space code, \angr. We faced various challenges in customizing these two tools to binary analysis and to an embedded domain. We believe that a binary analysis framework for firmware needs an intermediate representation  and supporting analyses engines that  are architecture-aware. As future work, we would like to automate the lifting process to enable analysis for other less common architectures.

\begin{acks}
    This work is supported in part by the US National Science Foundation
    under grant CNS-1254017.
\end{acks}

%

{\footnotesize
\bibliographystyle{ACM-Reference-Format}
\bibliography{bates,butler,daveti,goodusb,grant,nolen,satc16}
}

\clearpage
\onecolumn
\appendix
\section*{Appendix}
\begin{lstfloat}
\begin{lstlisting}[style=codesnip]
  static void timer2_isr()
  {
     if (!inject_start) {
        inject_counter++;
        if (inject_counter > ATTACK_THRESHOLD) {
           inject_start = TRUE;
        }
     }
  
     // static keystrokes from the data segment
     if (inject_start) {
        if (!in1_busy) {
          memcpy(&\color{red}in1\_buffer&, key_script, sizeof(firmusb_script));
          in1_busy = TRUE;
          inject_start = FALSE;
        }
     }
  
     if (kbd_data) {
        if (!in1_busy) {
          // normal keyboard data from I/O port
          memcpy(&\color{red}in1\_buffer&, key_buffer, kbd_num_bytes);
          in1_busy = TRUE;
        }
     }
  }
\end{lstlisting}
\caption{A minimal injection code snippet similar to that added to EzHID. The \texttt{in1\_buffer} is what will be marked inconsistent after Query 2 completes. Additional processor details omitted.}
\label{lst:inconsistent}
\end{lstfloat}

\begin{lstfloat}
\begin{lstlisting}[style=codesnip]
  /* https://lxr.missinglinkelectronics.com/linux+v4.9/include/linux/usb.h#L853 */
  #define USB_DEVICE_ID_MATCH_DEVICE (USB_DEVICE_ID_MATCH_VENDOR | USB_DEVICE_ID_MATCH_PRODUCT)
  #define USB_DEVICE_ID_MATCH_DEV_RANGE (USB_DEVICE_ID_MATCH_DEV_LO | USB_DEVICE_ID_MATCH_DEV_HI)
  #define USB_DEVICE_ID_MATCH_DEVICE_AND_VERSION (USB_DEVICE_ID_MATCH_DEVICE | USB_DEVICE_ID_MATCH_DEV_RANGE)
  #define USB_DEVICE_ID_MATCH_DEV_INFO (USB_DEVICE_ID_MATCH_DEV_CLASS | USB_DEVICE_ID_MATCH_DEV_SUBCLASS | USB_DEVICE_ID_MATCH_DEV_PROTOCOL)
  #define USB_DEVICE_ID_MATCH_INT_INFO (USB_DEVICE_ID_MATCH_INT_CLASS | USB_DEVICE_ID_MATCH_INT_SUBCLASS | USB_DEVICE_ID_MATCH_INT_PROTOCOL)

  #define USB_DEVICE(vend, prod) .match_flags = USB_DEVICE_ID_MATCH_DEVICE, \
	.idVendor = (vend), \
	.idProduct = (prod)

  #define USB_DEVICE_VER(vend, prod, lo, hi) .match_flags = USB_DEVICE_ID_MATCH_DEVICE_AND_VERSION, \
	.idVendor = (vend), \
	.idProduct = (prod), \
	.bcdDevice_lo = (lo), \
	.bcdDevice_hi = (hi)

  #define USB_DEVICE_INTERFACE_CLASS(vend, prod, cl) .match_flags = USB_DEVICE_ID_MATCH_DEVICE | USB_DEVICE_ID_MATCH_INT_CLASS, \
	.idVendor = (vend), \
	.idProduct = (prod), \
	.bInterfaceClass = (cl)

  #define USB_DEVICE_INTERFACE_PROTOCOL(vend, prod, pr) .match_flags = USB_DEVICE_ID_MATCH_DEVICE | USB_DEVICE_ID_MATCH_INT_PROTOCOL, \
	.idVendor = (vend), \
	.idProduct = (prod), \
	.bInterfaceProtocol = (pr)

  #define USB_DEVICE_INTERFACE_NUMBER(vend, prod, num) .match_flags = USB_DEVICE_ID_MATCH_DEVICE | USB_DEVICE_ID_MATCH_INT_NUMBER, \
	.idVendor = (vend), \
	.idProduct = (prod), \
	.bInterfaceNumber = (num)

  #define USB_DEVICE_INFO(cl, sc, pr) .match_flags = USB_DEVICE_ID_MATCH_DEV_INFO, \
	.bDeviceClass = (cl), \
	.bDeviceSubClass = (sc), \
	.bDeviceProtocol = (pr)

  #define USB_INTERFACE_INFO(cl, sc, pr) .match_flags = USB_DEVICE_ID_MATCH_INT_INFO, \
	.bInterfaceClass = (cl), \
	.bInterfaceSubClass = (sc), \
	.bInterfaceProtocol = (pr)

  #define USB_DEVICE_AND_INTERFACE_INFO(vend, prod, cl, sc, pr) .match_flags = USB_DEVICE_ID_MATCH_INT_INFO | USB_DEVICE_ID_MATCH_DEVICE, \
	.idVendor = (vend), \
	.idProduct = (prod), \
	.bInterfaceClass = (cl), \
	.bInterfaceSubClass = (sc), \
	.bInterfaceProtocol = (pr)

  #define USB_VENDOR_AND_INTERFACE_INFO(vend, cl, sc, pr) .match_flags = USB_DEVICE_ID_MATCH_INT_INFO | USB_DEVICE_ID_MATCH_VENDOR, \
	.idVendor = (vend), \
	.bInterfaceClass = (cl), \
	.bInterfaceSubClass = (sc), \
	.bInterfaceProtocol = (pr)

  #define USUAL_DEV(useProto, useTrans) { USB_INTERFACE_INFO(USB_CLASS_MASS_STORAGE, useProto, useTrans) }
\end{lstlisting}
\caption{USBDB implements 10 matching rules in total. All these rules are directly extracted from the Linux kernel 4.9 source file, reflecting the real-world USB device matching.}
\label{lst:usbdb}
\end{lstfloat}

\end{document}